\documentclass[aps,floats,nofootinbib,amssymb,preprint,superscriptaddress]{revtex4}
\usepackage{epsf,epsfig}

\usepackage{graphicx}

\newcommand{\be}{\begin{equation}}
\newcommand{\ee}{\end{equation}}
\newcommand{\bea}{\begin{eqnarray}}
\newcommand{\eea}{\end{eqnarray}}
\newcommand{\nn}{\nonumber}

\renewcommand{\check}{({\bf This statement needs to be checked!!})}

\newcommand{\half}{\frac{1}{2}}

\newcommand{\ba}{\begin{array}}
\newcommand{\ea}{\end{array}}
\newcommand{\bi}{\begin{itemize}}
\newcommand{\ei}{\end{itemize}}

\preprint{
\hbox to \hsize{
\hfill$\vcenter{\hbox{\bf MADPH-07-1492}
                \hbox{June 2007}}$}
}

\begin{document}
\title{\vspace*{.5in}
LHC Phenomenology of an Extended Standard Model with a Real Scalar Singlet}

\author{Vernon Barger}
\affiliation{Department of Physics, University of Wisconsin, Madison, WI 53706}

\author{Paul Langacker}
\affiliation{School of Natural Sciences, Institute for Advanced Study, Einstein Drive Princeton, NJ 08540}

\author{Mathew McCaskey}
\affiliation{Department of Physics, University of Wisconsin, Madison, WI 53706}
\author{Michael J. Ramsey-Musolf}
\affiliation{Department of Physics, University of Wisconsin, Madison, WI 53706}
\affiliation{California Institute of Technology, Pasadena, CA 91125}
\author{Gabe Shaughnessy}
\affiliation{Department of Physics, University of Wisconsin, Madison, WI 53706}

\thispagestyle{empty}
\begin{abstract}
\noindent Gauge singlet extensions of the Standard Model (SM) scalar sector may help remedy its theoretical and phenomenological shortcomings while solving outstanding problems in cosmology.  Depending on the symmetries of the scalar potential, such extensions may provide a viable candidate for the observed relic density of cold dark matter or a strong first order electroweak phase transition needed for electroweak baryogenesis. Using the simplest extension of the SM scalar sector with one real singlet field, we analyze the generic implications of a singlet-extended scalar sector for Higgs boson phenomenology at the Large Hadron Collider (LHC).  We consider two broad scenarios: one in which the neutral SM Higgs and singlet mix and the other in which no mixing occurs and the singlet can be a dark matter particle. For the first scenario, we analyze constraints from electroweak precision observables and their implications for LHC Higgs phenomenology. For models in which the singlet is stable, we determine the conditions under which it can yield the observed relic density,  compute the cross sections for direct detection in recoil experiments, and discuss the corresponding signatures at the LHC.   
\end{abstract}
\maketitle

\section{Introduction}

The minimal Standard Model (SM) of particle physics  agrees very well with precision measurements, and it provides a natural suppression of flavor changing neutral current effects as well as electric dipole moments  arising from electroweak CP-violation.  Despite its phenomenological success, however, the SM has well-known shortcomings, such as a large fine tuning required to obtain a Higgs mass that is not at the Planck scale for electroweak symmetry breaking at the TeV scale.   Fits to electroweak precision data also suggest a light Higgs boson, in mild conflict with the excluded region of masses set by LEP, though this tension can be relieved with a light Higgs boson somewhat above the LEP limit of 114 GeV~\cite{Barate:2003sz}.  The SM also fails to provide a particle physics explanation for cosmological observations, such as the predominance of visible matter over antimatter and the non-luminous dark matter (DM) whose contribution to the cosmic energy density is about five times larger than that of the visible matter.  Indeed, the abundances of both visible and dark matter  -- along with neutrino oscillations -- are the most direct evidence for physics beyond the SM.

A simple extension of the SM that can help solve these problems is the addition of a singlet scalar field.  Singlet extensions have been studied in the SM \cite{McDonald:1993ex,Bento:2000ah,Burgess:2000yq,Davoudiasl:2004be,Schabinger:2005ei,O'Connell:2006wi,Kusenko:2006rh,Bahat-Treidel:2006kx} and in supersymmetry \cite{Barger:2006dh,Barger:2006sk,Barger:2007nv,Barger:2007ay,Ellis:1988er,NMSSM1,NMSSM2,nMSSM, Panagiotakopoulos:2000wp,Menon:2004wv,Dedes:2000jp,UMSSM1,umssm2,deCarlos:1997yv,smssm,Han:2004yd,Choi:2006fz,Cheung:2007sv,King:2005my,King:2005jy,Li:2006xb,Lopez-Fogliani:2005yw,Schuster:2005py}. With the imminent operation of the Large Hadron Collider (LHC), it is worthwhile considering the implications of singlet extensions of the scalar sector of the SM for  Higgs boson studies at the LHC. In this paper, we delineate the broad outlines of Higgs phenomenology at the LHC in these singlet-extended scenarios, addressing the following questions: (1) To what extent can the presence of additional singlet scalars that mix with the SM Higgs boson affect the Higgs discovery potential at the LHC? (2) If a neutral scalar is discovered at the LHC, to what extent can one determine that it is a \lq\lq SM-like" or \lq\lq singlet-like" scalar?  (3) If the singlet is stable and provides for the observed relic abundance of cold dark matter, how will it affect Higgs boson searches at the LHC?  (4) What are the direct detection prospects in elastic scattering experiments of an augmented scalar sector that includes a stable singlet?

To address these issues, we consider the simplest extension of the SM scalar sector that involves the addition of a real scalar singlet field $S$ to the SM Lagrangian. Although it is possible to generalize to scenarios with more than one singlet, the simplest case of a single additional singlet scalar provides a useful framework for analyzing the generic implications of an augmented scalar sector for LHC phenomenology\footnote{We note, however, that the authors of Ref.~\cite{Bahat-Treidel:2006kx} observed that the presence of a very large number ($\gg 10$) of light scalars could degrade the Higgs discovery potential at the LHC. Here, we concentrate on the opposite extreme $N\ll 10$ that seems well-motivated theoretically.}. Following the notation of Ref. \cite{O'Connell:2006wi}, the most general Higgs potential with one additional singlet field that does not obtain a vacuum expectation value (VEV) is
\bea
V &=& {m^2\over 2} H^\dagger H+{\lambda\over 4}(H^\dagger H)^2+{\delta_1\over 2} H^\dagger H S+{\delta_2\over 2} H^\dagger H S^2\nn\\
&+&\left({\delta_1 m^2\over 2 \lambda}\right) S + {\kappa_2 \over 2} S^2+{\kappa_3\over 3} S^3+{\kappa_4 \over 4} S^4,
\label{eqn:hpot}
\eea
where $H$ is the $SU(2)$ double field and where $m^2$ and $\lambda$ are the usual SM parameters of the Higgs potential.  Combining this singlet extended Higgs sector with the rest of the SM gives the \lq\lq xSM", the extended Standard Model.  The coefficient of the linear term in $S$  is chosen so that $S$ does not acquire a VEV (or equivalently, emerges after shifting $S$ to remove its VEV). The parameter $\delta_1$  appearing there and in the $H^\dag HS$ term governs the degree of mixing between the $S$ and the SM Higgs, $h$. In the absence of such mixing, the singlet mass is determined by $\kappa_2$ and $\delta_2$.

\subsection{Higgs Mixing Case}

In general, the singlet field, $S$, mixes with the SM Higgs boson, $h$, allowing it to couple to the same states as the SM Higgs. As noted in Ref.~\cite{O'Connell:2006wi}, the decay branching ratios for the lightest of the two neutral scalars, $H_1$, will be identical to those of a SM Higgs boson having the same mass, while its production rate will be reduced from that of a SM Higgs by the square of the mixing parameter. If the mass of the second scalar, $H_2$,  is more than twice that of the first, its branching ratios to conventional SM Higgs decay products will be reduced because the decay $H_2\to 2H_1$ becomes kinematically allowed.  Moreover, the presence of this \lq\lq Higgs splitting" channel would result in exotic Higgs decay products, such as four $b$-jets or $b{\bar b}\tau^+\tau^-$. This channel is particularly interesting if the $H_2$ is singlet-like, since it would be a signature of singlet extensions that provide for a strong, first order electroweak phase transition (EWPT) as needed to explain the baryon asymmetry of the universe (BAU)\cite{Profumo:2007wc}.  Note that while a singlet that mixes with the SM Higgs is unstable and, therefore, not a candidate for particle dark matter, the abundance of dark matter may be explained by the QCD axion \cite{Sikivie:2006ni}.

In what follows, we analyze the consequences for LHC Higgs phenomenology from the general features of Eq.~(\ref{eqn:hpot}). In doing so, we discuss the implications of Higgs boson searches at LEP and of electroweak precision observables (EWPO). Both considerations lead to rather severe restrictions on the phenomenologically relevant parameters in the potential. Indeed, a key feature of our analysis involves the tension between EWPO, which favor light scalars that have a significant $SU(2)$ component, and the LEP searches, which allow scalars with masses below $114$ GeV only if they have a relatively small $SU(2)$ admixture. The presence of the augmented scalar sector in the xSM relaxes this tension compared to the situation in the SM, where EWPO favor a Higgs mass below 150 GeV \cite{Kile:2007ts}\footnote{This bound pertains to the impact of EWPO alone an does not incorporate the constraints from the LEP 2 direct search lower bound. Including the latter can increase the upper limit by $\sim 40$ GeV \cite{LEPEWWG:2007}.}. When the mixing between $h$ and $S$ is maximal and each mass eigenstate couples to SM gauge and matter fields with equal strength, the mass of the heavier scalar can be as large as $\sim 220$ GeV (see Figure 10 of Ref. \cite{Profumo:2007wc}.). Away from maximal mixing, the upper bound on the SM-like scalar becomes smaller, while that of the singlet-like scalar can be larger. Even with these relaxed EWPO restrictions on the xSM scalar masses, the competing considerations from EWPO and direct Higgs searches strongly affect the character of possible models leading to  discovery prospects of neutral scalars at the LHC. 

In this context, it is important to bear in mind that an extended scalar sector described by Eq.~(\ref{eqn:hpot}) may  not be the only manifestation of the larger framework in which the Standard Model is embedded. The presence of additional new physics that significantly affects EWPO will modify the analysis of the foregoing considerations \cite{Peskin:2001rw} and could allow heavier scalars with significant $SU(2)$ fraction\footnote{The figures in this paper show both the cases in which the EWPO constraints are imposed and those in which they are not, both for comparison and to allow for the possibility of compensating new physics.}. 

In the absence of compensating new physics, we find that if a neutral scalar is discovered at the LHC with mass above $\sim 160$ GeV, then it is quite likely to be the heavier scalar ($H_2$) and to contain a significant mixture of the $S$ with the $h$. The singlet admixture must be large enough to suppress the effect of the $H_2$ on electroweak radiative corrections but still small enough to allow significant coupling to conventional SM Higgs decay modes needed for  its discovery. Importantly, the possibility of discovering an EWPO-compatible neutral scalar in this mass range is one consequence of an augmented scalar sector and would provide strong evidence for physics beyond the minimal SM.
Conversely, if the mass of the scalar is lighter than $\sim 160$ GeV, then it is most likely to be a SM-like $H_1$.  In either case,  if such a scalar discovery is made at the LHC then it is possible to determine or limit the degree of $S$-$h$ mixing by observing the event rate of the $H_j\to ZZ\to 4\ell$ channel  or $H_j\to WW\to \ell\nu jj$, which is feasible with $\geq30$ fb$^{-1}$ of data if the $SU(2)$ fraction of $H_j$ is $\gtrsim 40\%$.

It is also possible that the presence of an augmented scalar sector would reduce the probability of making a $5\sigma$ discovery using conventional SM Higgs search channels. Nevertheless, a neutral scalar that does not yield such a $5\sigma$ discovery may be observed if the Higgs to two Higgs splitting process is kinematically allowed.  In this case, if the $H_2$ is SM-like, then EWPO imply that it must be lighter than about $160$ GeV, and if the Higgs splitting channel is open then it could yield a significant number of four $b$-jet events.  Conversely, if the $H_2$ is singlet-like, then its mass must be $\gtrsim 230$ GeV, since the pair of $H_1$ into which it decays must each be SM-like and have mass $> 114$ GeV to satisfy the LEP search limit.  For some models, one could observe several hundred $b{\bar b}b{\bar b}$ jet events that reconstruct to the $H_2$ mass (before cuts) with 30 fb$^{-1}$ integrated luminosity, with the number of such events decreasing with the mass of the $H_2$.  However, it is quite possible that the Higgs splitting decays of singlet-like $H_2$ would not produce enough exotic final state events above backgrounds to be conclusively identified at the LHC. In this case, one would look to future Higgs studies at a Linear Collider. 

\subsection{Dark matter singlet}

If the potential in Eq.~(\ref{eqn:hpot}) displays a $\mathbb{Z}_2$ symmetry ($\delta_1=\kappa_3=0$), vertices involving an odd number of singlets field do not exist, making the singlet stable.  In this case, the singlet can be a viable candidate for particle dark matter (DM).  The parameter space of this model is then constrained to accommodate the density of relic DM particles in the universe implied by the cosmic microwave background and other astrophysical observations \cite{Spergel:2006hy,Yao:2006px}.  

Scalar singlet DM in this model can also have a significant impact for Higgs boson searches at the LHC  since the singlet only couples to SM Higgs decay modes via the Higgs boson.  In many cases, the Higgs boson can decay to two singlets, thereby reducing the likelihood of discovering the Higgs boson in traditional search modes.  However, the Higgs can still be discovered with a search via weak boson fusion and $Z$-Higgstrahlung where the Higgs decays to states that are invisible \cite{Eboli:2000ze,Davoudiasl:2004aj}.  Other, indirect, non-accelerator avenues toward observing the stable singlet are through dark matter recoil detection experiments.  Within the parameter space we consider, the singlet is expected to generate a proton spin-independent cross section in the range of $10^{-8} - 10^{-9}$ pb for such recoil experiments.  A cross-section of this order is within the range of upcoming indirect experiments \cite{DMSAG}. The $S$ has only scalar interactions with matter, so it would not generate a spin-dependent signal in future DM searches.

The analysis leading to these (and other conclusions) is organized as follows: In Section \ref{sect:higgs}, we discuss singlet mixing in the Higgs sector and the considerations from LEP Higgs searches.  In Section \ref{sect:ewp}, we discuss the constraints from EWPO on the scalar sector.  The resulting implications for LHC Higgs studies are discussed in Section \ref{sect:statsig}. The singlet as a dark matter candidate is discussed in Section \ref{sect:dm} where predictions for direct detection rates for spin-independent scattering on nucleons are given.  We summarize our main results with model illustrations in Section \ref{sect:illust} and conclude in Section \ref{sect:concl}. Technical details regarding scalar contributions to gauge boson propagators as needed for the analysis of EWPO appear in the Appendix.

\section{The Singlet as an Extra Higgs boson}
\label{sect:higgs}

Singlet mixing in the Higgs sector is a well-studied effect in the xSM \cite{Davoudiasl:2004be,O'Connell:2006wi} and MSSM \cite{Barger:2006dh,Barger:2006sk} as well as radion mixing in Randall-Sundrum models \cite{Kribs:2001ic}.

The mass-squared matrix of the Higgs sector in the singlet extended SM is
\be
M_H^2 = \left(
\begin{array}{cc}
\lambda v^2/2 & \delta_1 v/2\\
\delta_1 v/2 & \lambda_S v^2/2\\
\end{array}
\right),
\ee
where $v=\sqrt{2} \langle H^0\rangle = 246$ GeV and $\lambda_S \equiv \delta_2 + 2\kappa_2 /v^2$.  This matrix can be diagonalized by a rotation matrix $R_{ij}$ to obtain the mass eigenstates
\be
\label{eq:masseigenstates}
\left(
\begin{array}{c}
H_1\\
H_2   \\
\end{array}
\right)
=  \left(
\begin{array}{cc}
\cos\phi & \sin\phi\\
-\sin\phi & \cos\phi\\
\end{array}
\right) \left(
\begin{array}{c}
h\\
S  \\
\end{array}
\right),
\ee
The masses and mixing angles are given by
\bea
M^{2}_{1,2}&=&\frac{(\lambda+\lambda_{S})v^{2}}{4} \mp
\frac{\sqrt{v^{4}(\lambda-\lambda_{S})^{2}+4v^{2}\delta_{1}^{2}}}{4}
\nonumber\\
\label{eq:massmix}
\tan\phi&=&\frac{2\delta_{1}}{v(\lambda-\lambda_{S})-\sqrt{v^{2}(\lambda-\lambda_{S})^{2}+4\delta_{1}}}
\eea
where the states are ordered according to their mass: $M_{H_1} < M_{H_2}$.  (We note that these conventions differ from those of Refs. \cite{O'Connell:2006wi,Profumo:2007wc},  where the eigenstates denote the $SU(2)$-like and singlet-like scalars, respectively).

Singlet mixing reduces the coupling strengths of the Higgs state, $H_i$, to all SM fields by the factors
\be
g^2_{H_1} = \cos ^2\phi, \quad g^2_{H_2} = \sin^2\phi.
\label{eq:coup}
\ee 
Since the reductions are universal for all SM interactions, the branching fractions of the Higgs bosons to the SM modes do not differ from those of the SM if additional decay modes are not accessible.  Therefore, the lightest Higgs boson often obeys the LEP lower limit of 114 GeV.  Exceptions are possible if $\cos^2\phi$ is small, reducing the $ZZh$ coupling or if the Higgs is very light, below the threshold for decays to $b \bar b$.  This is similar to the case of a light $A_1$ in the NMSSM \cite{Dermisek:2006wr,Dermisek:2005gg}.

If kinematically allowed, the heavier Higgs boson may decay to pairs of the light Higgs, altering the $H_2$ branching fractions to SM modes ($X_{SM}$)
\be
\label{eqn:bfreduction}
\text{BF}(H_2 \to X_{SM}) ={g_{H_2}^2 \text{BF}(h_{SM} \to X_{SM}) \Gamma_{h_{SM}}\over g_{H_2}^2 \Gamma_{h_{SM}}+\Gamma({H_2 \to H_1 H_1})}.
\ee
Here the heavy Higgs decay rate is given by
\be
\Gamma(H_2\to H_1 H_1) = {|g_{211}|^2\over 32 \pi M_{H_2}} \sqrt{1-{4 M_{H_1}^2\over M_{H_2}^2}},
\ee
with the $H_2 H_1 H_1$ coupling given by
\bea
g_{211}&=& \half \delta_1 \cos^3\phi+ (2 \kappa_3 - \delta_1)\sin^2\phi \cos\phi\nn\\
&&-  {v\over 2} \sin\phi \left( 3 \lambda \cos^2\phi+2\delta_2(1-3\cos^2\phi)\right)
\eea
This decay is accessible only if 
\be
(\lambda - 4\lambda_S)(4\lambda - \lambda_S) +25 {\delta^2_1\over v^2}>0.
\ee

A reduction in the branching fractions and coupling can result in a decrease in the SM statistical significance of Higgs discovery at the LHC.  This reduction factor can be written as a product of the production strength and the branching fractions relative to the SM
\bea
\label{eqn:xi2def}
\xi_{i}^2 &=& g_{H_i}^2 {\text{BF}(H_{i} \to X_{SM})\over\text{BF}(h_{SM} \to X_{SM}) }\nn\\
&=&\Bigg\{ \begin{array}{ccc}g_{H_1}^2& &  i=1\\
g_{H_2}^2 {g_{H_2}^2 \Gamma_{h_{SM}}\over g_{H_2}^2 \Gamma_{h_{SM}}+\Gamma({H_2 \to H_1 H_1})} & &i=2
\end{array}
\eea
To illustrate these effects, we have performed a numerical scan over the parameter space describing the extended  scalar sector of the xSM.  The ranges of values we adopt are 
\be
\begin{array}{ccccc}
0&\le& \lambda&\le& 3\\
-3&\le& \delta_2& \le& 3\\
-200\text{ GeV}&\le& \delta_1& \le& 200\text{ GeV}\\
-(500 \text{ GeV})^2&\le&\kappa_2&\le&(500 \text{ GeV})^2\\
-1000\text{ GeV}&\le& \kappa_3&\le& 1000\text{ GeV},
\label{eq:ranges}
\end{array}
\ee
where the quadratic scalar parameter, $\kappa_2$, is absorbed into $\lambda_S$ as above.  The quartic parameter, $\kappa_4$, does not affect the Higgs sector since $\langle S\rangle = 0$.  The range of  parameters scanned is sufficiently large to show the varied states that can exist at the few hundred GeV mass scale.  The ranges on the dimensionless parameters $\lambda$ and $\lambda_S$ are broadly consistent with considerations of perturbativity \cite{Arnold:1992rz} (see also Ref.~\cite{Profumo:2007wc}). As we discuss below, EWPO restrict the mass of a SM-like scalar to be well below $v$, leading to reduced ranges for $\lambda$ and $\lambda_S$. 

\begin{figure}[t]
\begin{center}
\includegraphics[angle=-90,width=0.49\textwidth]{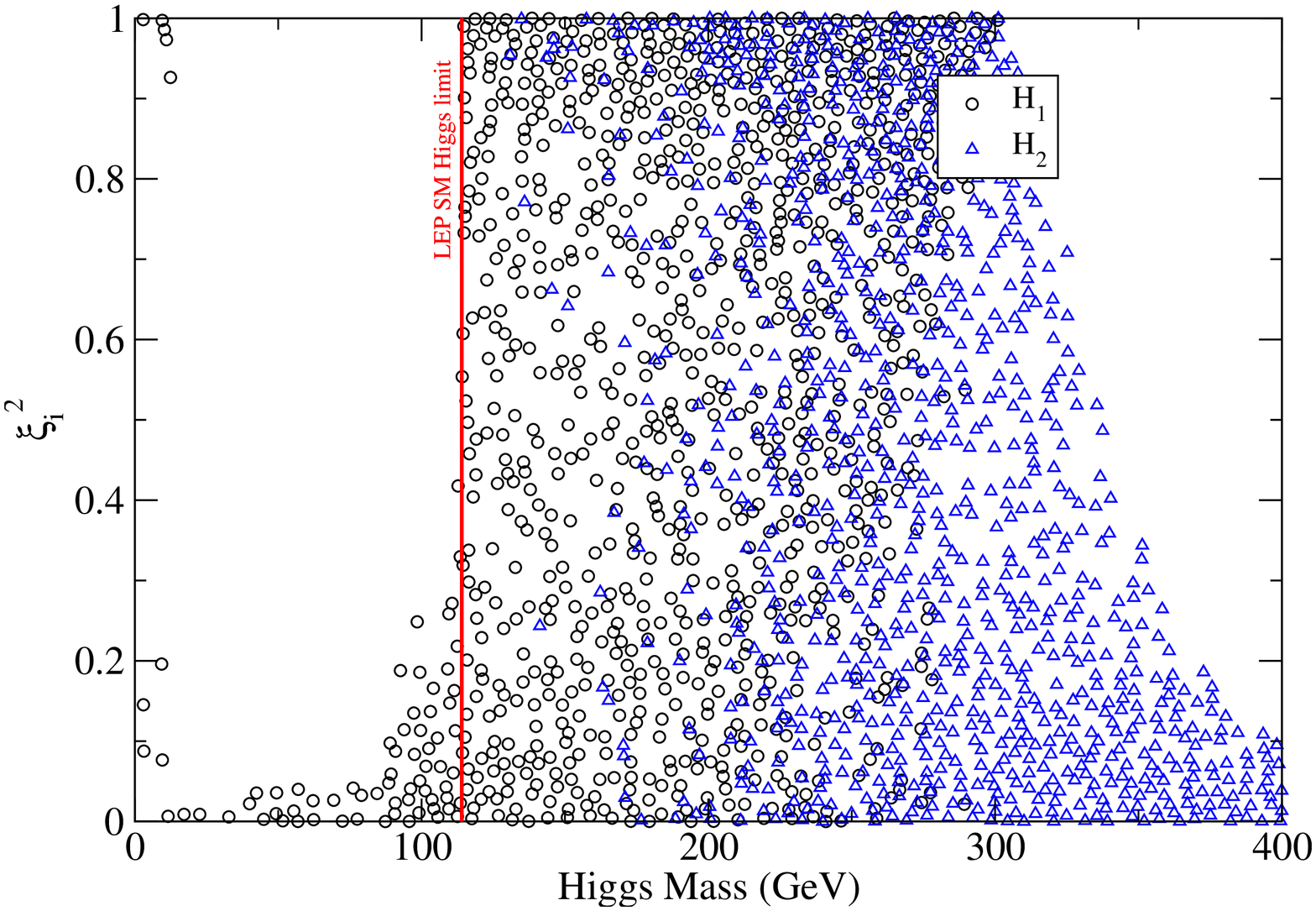}
\includegraphics[angle=-90,width=0.49\textwidth]{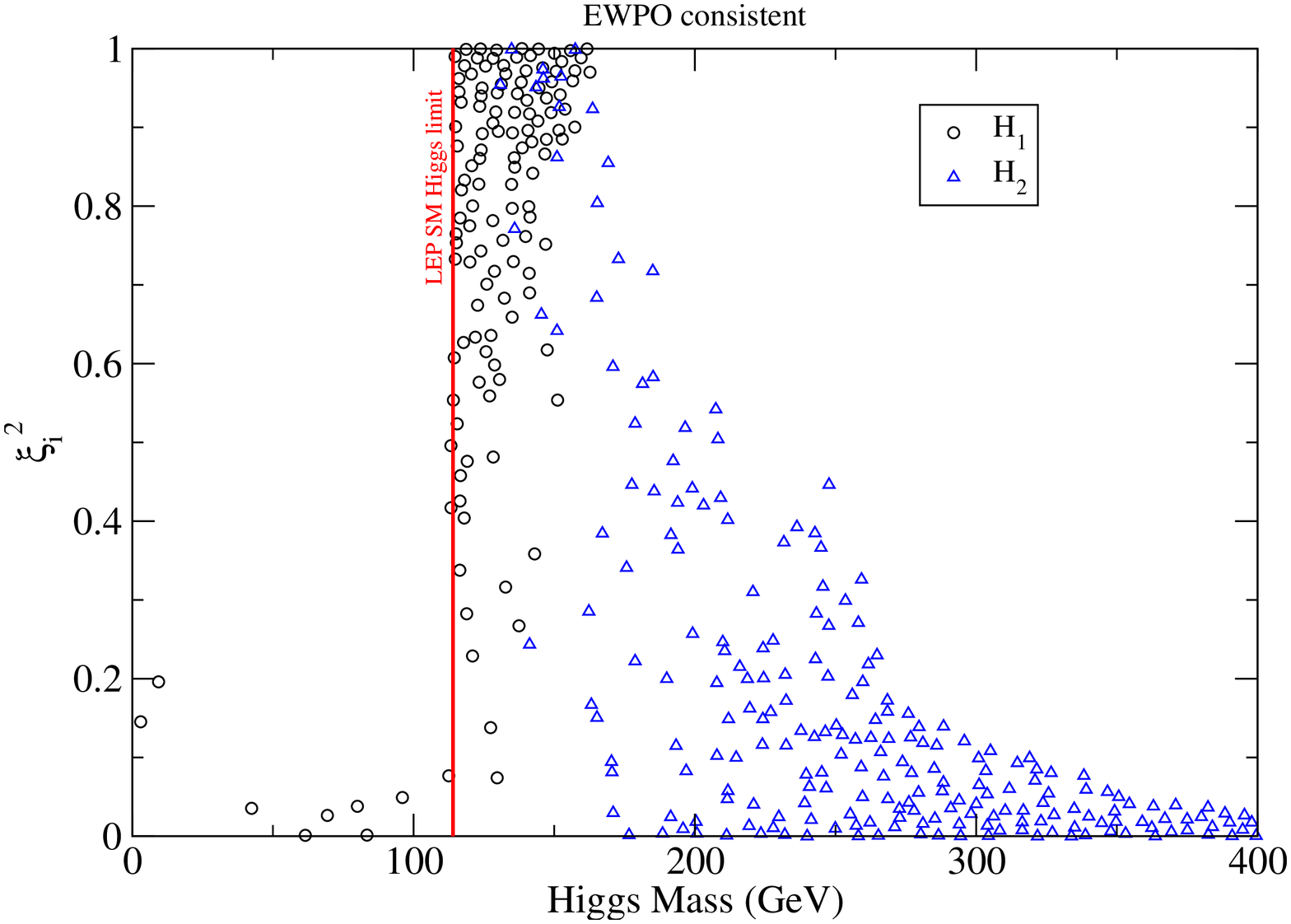}
\caption{Reduction factor of the Higgs boson signal with respect to the SM.  The LEP SM Higgs mass bound for $\xi_i^2=1$ is marked as the vertical red line.  All of the points indicated are consistent with the LEP bounds. The right (left) panel does (not) include constraints from EWPO as discussed below. }
\label{fig:sigreduce}
\end{center}
\end{figure}

Fig. \ref{fig:sigreduce} shows the reduction factor, $\xi_i^2$, of the signal compared to that of the SM Higgs.  When the coupling $\delta_1 v$ is large, mixing is strong enough ({\em e.g.}, $\cos^2\phi$ is sufficiently suppressed) to permit the lightest Higgs ($H_1$) to be below the LEP limit shown by the vertical line.  Alternatively, if $\delta_1 v$ is small then the lightest Higgs can be dominantly singlet if $\lambda_S \lesssim 0.45 \lesssim \lambda$. In either case, the Higgs can be lighter than the LEP limit due to its weak coupling to SM fields.  For a very light Higgs, the amount of mixing can be severely limited by experimental limits on $B \to H_i X$ and $\Upsilon\to H_i \gamma$ decays \cite{O'Connell:2006wi,Dermisek:2006py}.  

\section{Electroweak Precision Constraints}
\label{sect:ewp}

\begin{figure}[t]
\begin{center}
\includegraphics[width=0.79\textwidth]{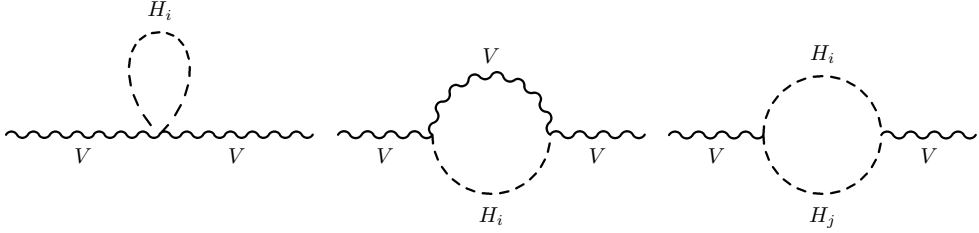}
\caption{Feynman diagrams of gauge boson propagators that are affected by Higgs bosons.}
\label{fig:ewfd}
\end{center}
\end{figure}

The mixing of the neutral $SU(2)$ and singlet scalars will affect electroweak precision observables (EWPO) through changes in the gauge boson propagators. To analyze the corresponding implications for the xSM we have computed the scalar contributions to the gauge boson propagators generated by the one-loop diagrams of  Fig. \ref{fig:ewfd}. In the presence of an additional neutral scalar, only the scalar contributions to the $W$ and $Z$-boson propagator functions, $\Pi_{WW}(q^2)$ and $\Pi_{ZZ}(q^2)$, respectively, differ from those in the SM. Since the neutral scalars have no electromagnetic coupling, $\Pi_{\gamma\gamma}(q^2)$ and $\Pi_{Z\gamma}(q^2)$ are unaffected. We work in the Feynman gauge and employ $\overline{MS}$ renormalization. In order to delineate the various contributions, it is useful to relate the neutral mass eigenstates to components of the neutral $SU(2)$ and singlet scalar field {\em via}
\be
\left(
\begin{array}{c}
h \\
\xi\\
S
\end{array}\right) = \mathbb{V}_H
\left(
\begin{array}{c}
H_1\\
H_2\\
G^0
\end{array}\right)
\ee
where the $H_j$ are the mass eigenstates given in Eq.~(\ref{eq:masseigenstates}), $G^0$ is the neutral would-be Goldstone boson, and the neutral component of the $SU(2)$ scalar is given by 
$H^0=(v+h+i\xi)/\sqrt{2}$. The  mixing matrix $\mathbb{V}_H$ is given by
\be
\mathbb{V}_H=\left(
\begin{array}{ccc}
\cos\phi & -\sin\phi & 0 \\
0 & 0 & 1\\
\sin\phi & \cos\phi & 0 
\end{array}
\right) \ \ \ .
\ee

Explicit expressions for the modified scalar contributions to the $W$ and $Z$ propagator functions  for an arbitrary number of additional real (neutral) scalars computed in  the $\overline{MS}$ scheme are given in Appendix \ref{apx:prop}. The corresponding $\overline{MS}$ renormalized quantities are denoted ${\hat\Pi}_{VV}(q^2,\mu)$, where $V=W$ or $Z$ and $\mu$ is the renormalization scale. Here, we take $\mu=M_Z$. The scalar contributions to the ${\hat \Pi}_{VV}$ are governed by three parameters: the two masses, $M_{H_{1,2}}$, and the mixing angle, $\phi$. Constraints on these quantities implied by EWPO translate into restrictions on the parameters appearing in the potential via Eq.~(\ref{eq:massmix}).

The effect of scalar contributions to the ${\hat\Pi}_{VV}$ on $Z^0$-pole observables and the $W$-boson mass can be characterized by the oblique parameters $S$, $T$, $U$, and $V$. The impact on observables associated with other processes, such as atomic parity violation \cite{Wood:1997zq} and parity-violating electron scattering asymmetries \cite{Anthony:2003ub} at low momentum transfer, require inclusion of the additional oblique parameter $X$ that compares the   the $Z$-$\gamma$ mixing tensor, ${\hat\Pi}_{Z\gamma}$, at $q^2=M_Z^2$ and $q^2=0$. However, neutral scalars do not contribute to ${\hat\Pi}_{Z\gamma}$ at one-loop order, so their impact on low-energy precision observables is characterized solely by $S$ and $T$ (non-oblique contributions to low-energy precision observables are suppressed by the light fermion Yukawa couplings). When the mass scale of new particles that contribute to the ${\hat\Pi}_{VV}$ is well above the electroweak scale, inclusion of only $S$, $T$, and $U$ provides a good approximation to the exact contributions from these particles to both low- and high-energy EWPO. In the present instance, the masses of the neutral scalars can be close to the electroweak scale, so this approximation is not {\em a priori} valid. In particular, the $V$ parameter that enters the $Z$ partial widths and that depends explicitly on the first derivative of ${\hat\Pi}_{ZZ}(M_Z^2)$ may not be entirely negligible. Nevertheless, one may obtain a reasonable picture of the implications of EWPO for the xSM by first considering the leading terms in the derivative expansion characterized by $S$, $T$, and $U$.

To that end, we have performed a fit to EWPO using the GAPP routine \cite{Erler:1999ug}  assuming only SM contributions to the various amplitudes and extract values of $S$, $T$, and $U$ for fixed values of the SM Higgs mass, $m_H=114.4$ GeV~\cite{Barate:2003sz} and a top quark mass $m_t=170.9\pm 1.8$ GeV ~\cite{cdfcollab:2007bx}. The observables included in our fit include those entering the global analysis given in the Review of Particle Properties \cite{Yao:2006px} and encompass $Z$-pole precision measurements, the $W$-boson mass, and several low-energy observables including atomic parity violation \cite{Wood:1997zq}  and parity-violating M\o ller scattering \cite{Anthony:2003ub}.  As discussed in Ref.~\cite{Profumo:2007wc}, the results of the fit can be interpreted as constraints on $\Delta{\cal O}_j\equiv{\cal O}_j^{\rm xSM}-{\cal O}_j^{SM}$, where ${\cal O}_j$ is any one of the three leading order oblique parameters. The ${\cal O}_j^{SM}$ are the contributions to the oblique parameters from a single $SU(2)$ neutral scalar with mass $114.4$ GeV. These contributions are included in the GAPP routine, so they must be subtracted out when using the results of the fit to analyze an augmented scalar sector that includes the SM Higgs. The central values of the oblique parameters from the global EW fit are
\bea
\nonumber
T^{\rm xSM}-T^{\rm SM} &=& -0.111\pm 0.109\\
\label{eq:STUbest}
S^{\rm xSM}-S^{\rm SM} &=& -0.126\pm 0.096\\
\nonumber
U^{\rm xSM}-U^{\rm SM} &=& 0.164\pm 0.115
\eea
and the covariance matrix is
\be
\sigma_{ij}^2 = \left(\begin{array}{ccc}
 1 & 0.866 & -0.588\\
 0.866 & 1 & -0.392\\
 -0.588 & -0.392 & 1\\
\end{array}
\right)
\ee
From these parameters, a value of $\Delta \chi^2$ can be found:
\be
\Delta \chi^2 = \sum_{i,j}\left( \Delta{\cal O}_i -  \Delta{\cal O}^0_i\right) (\sigma^2)_{ij}^{-1} \left( \Delta{\cal O}_j -  \Delta{\cal O}^0_j\right).
\ee
The 95\% C.L. ellipsoid for the  $\Delta {\cal O}_i$ is obtained  by requiring $\Delta \chi^2 < 7.815$.  We note that the central values for $T$ and $S$  in Eq.~(\ref{eq:STUbest}) lie somewhat further from zero than those obtained by the LEP Electroweak Working Group \cite{LEPEWWG:2007}, whose fit includes only the high-energy precision observables. Our central values and ranges, however, are consistent with those given in the PDG \cite{Yao:2006px}, but slightly shifted due to use of a more recent value for $m_t$.

To understand the implications of our fit for the extended scalar sector, it is instructive to consider scalar contributions to the  $T$ parameter, for which a simple analytic expression obtains
\begin{eqnarray}
\label{eq:Txsm}
-T^{\rm xSM}&=&  \frac{1}{8\pi s_W^2}\Bigl\{ 
\sum_{a,b}\, \Bigl[(V_H^{1a})^2(V_H^{2b})^2-(V_H^{1a}V_H^{2a})(V_H^{1b}V_H^{2b})\Bigr]  {\tilde F} ({\bar m}_a^2,{\bar m}_b^2)\\ \nonumber
&+&2\sum_a\, (V_H^{1a})^2\left[ \frac{1}{c_W^2}F_0(1/c_W^2, {\bar m}_a^2,0)- F_0(1, {\bar m}_a^2,0)\right] - \sum_{j=1,2; a} (V_H^{ja})^2\, {\tilde F}({\bar m}_a^2,1)\Bigr\}\nonumber
\end{eqnarray}
where ${\bar m}^2\equiv m^2/M_W^2$, $c_W^2=M_W^2/M_Z^2$, 
\begin{eqnarray}
\nonumber
{\tilde F}(m_a^2, m_b^2) & = & \frac{1}{4}\left(m_a^2+m_b^2\right)-\frac{m_a^2 m_b^2}{2(m_a^2-m_b^2)}\, \ln\frac{m_a^2}{m_b^2}\\
\nonumber
F_n(m_a^2, m_b^2, q^2) & = & \int_0^1\, dx\, x^n\, \ln \Bigl[(1-x)m_a^2+x m_b^2 -x(1-x)q^2\Bigr]
\end{eqnarray}
with
\begin{equation}
F_0(m_a^2, m_b^2,0) = \frac{1}{(m_a^2-m_b^2)}\, \left[m_a^2\ln m_a^2-m_b^2\, \ln m_b^2\right]-1\ \ \ .
\end{equation}
Here the index \lq\lq $a$" runs over all mass eigenstates ($a=1,2,3$), including the would be Goldstone bosons\footnote{Recall that in the $R_\xi$ gauge, the masses appearing in the would be Goldstone propagators are those of the corresponding massive gauge bosons.  Therefore, $m_3 = M_Z$.}. The analogous expression for $S^{\rm xSM}$ and $U^{\rm xSM}$ can be obtained using the results for the $\Pi_{WW}(q^2)$ and $\Pi_{ZZ}(q^2)$  given in the Appendix.  The expression in Eq.~(\ref{eq:Txsm}) accommodates the possibility of adding more than one real scalar to the theory. Increasing the number of real scalar fields simply increases the size of the mixing matrix involving these scalars and the real part ($h$) of the neutral $SU(2)$ field. In this case, the matrix element $V_H^{2a}$ is non-vanishing only for the would-be Goldstone boson, the cross terms involving $V_H^{1a} V_H^{2a}$ {\em etc.} vanish, and the intermediate states involve only a single physical scalar or one physical scalar and one would-be Goldstone boson.

For the present case of a single real scalar singlet, we obtain
\begin{eqnarray}
\nonumber
 \label{eq:Txsm1}
T^{\rm xSM} & = & -\left(\frac{3}{16\pi{\hat s}^2}\right)\Biggl\{
 \cos^2\phi\, \Bigl[\frac{1}{c_W^2}\left(\frac{m_1^2}{m_1^2-M_Z^2}\right)\, \ln\, \frac{m_1^2}{M_Z^2} -\left(\frac{m_1^2}{m_1^2-M_W^2}\right)\, \ln\, \frac{m_1^2}{M_W^2}\Bigr]\\ 
&&+ \sin^2\phi\, \Bigl[\frac{1}{c_W^2}\left(\frac{m_2^2}{m_2^2-M_Z^2}\right)\, \ln\, \frac{m_2^2}{M_Z^2} - \left(\frac{m_2^2}{m_2^2-M_W^2}\right)\, \ln\, \frac{m_2^2}{M_W^2}\Bigr]\Biggr\}+\cdots\ \ .
\end{eqnarray}
where the \lq\lq $+\cdots$" denote terms that have no dependence on the mass and mixing of the Higgs states and that cancel from the quantity $\Delta T=T^{\rm xSM}-T^{\rm SM}$ that is constrained by our fit.  The corresponding expression in the SM corresponds to taking $\cos\phi=1$ and $m_1\to m_h$. From Eq.~(\ref{eq:Txsm1}), we observe that for very heavy scalars ($m_i \gg M_Z$), the dependence of $T$ on the scalar masses is logarithmic, but for lighter scalars, the $m_i$-dependence is more complicated. The contributions from the two mass eigenstates are identical in form but weighted by the appropriate factors of $\cos^2\phi$ and $\sin^2\phi$. Similar observations apply to the scalar contributions to $S^{\rm xSM}$ and $U^{\rm xSM}$. 

It is well known that for the SM, EWPO favor a light Higgs, with the value for $m_h$ that minimizes the $\chi^2$ falling below the direct search lower bound.  We would, thus,  expect that EWPO favor a relatively light Higgs that is dominantly SM-like while there are less restrictions on the singlet-like state.  These expectations are born out by the results of our fit, which have been reported in more detail in Ref.~\cite{Profumo:2007wc}. In that work, the dependence of $\Delta\chi^2$ on $m_i\equiv M_{H_i}$ was studied. When the mass of the heaviest scalar, $H_2$, is large and the mixing angle small, the dependence of $\Delta\chi^2$ on the mass of the lighter scalar $H_1$ is close to that for a single, pure $SU(2)$ Higgs. In this limit, the EWPO constraints have minimal implications for the properties of the $H_2$. With increasing mixing angle, however, the EWPO favor increasingly lighter $H_2$ as well as a relatively light $H_1$. For maximal mixing, $m_2\lesssim 220$ GeV at 95\% C.L. (see Fig. 10 of Ref.~\cite{Profumo:2007wc}.).  We note in passing that these trends follow largely from the EWPO constraints on $S$ and $T$, as the scalar contributions to $U$ tend to be rather small, and that a fit that includes only $S$ and $T$ may yield somewhat different constraints on the parameters of the extended scalar sector than with $U=0$ imposed (see, {\em e.g.}, Ref.~\cite{Yao:2006px}). We also reiterate that inclusion of the direct search lower bounds in the fit may relax the upper limits on the scalar masses, as occurs in the SM. We defer a detailed treatment of this possibility to a future analysis. 

In the remainder of this study, we will indicate the impact of EWPO constraints on various aspects of singlet Higgs phenomenology. To that end, we show -- in the right panel of Fig. \ref{fig:sigreduce} -- the impact of the EWPO constraints on the signal reduction as a function of $M_{H_i}$. Models with $M_{H_i}\gtrsim 220$ GeV and $\xi_i^2\gtrsim 0.5$ are excluded by the precision electroweak data. Most of the surviving models have  a light scalar that is strongly SM-like and a heavy scalar that mixes very weakly with the neutral $SU(2)$ scalar. It is possible, however, to realize models with moderately heavy scalars $\sim 200-300$ GeV and moderate mixing, $\xi_i^2\sim 0.2-0.4$, and there are a few points with a light singlet-like scalar.  In this respect, the implications of our fit for the mass and couplings of the singlet-like scalar differ from those of Ref.~\cite{Bahat-Treidel:2006kx}, which considered additional singlet scalars that mix with the neutral $SU(2)$ scalar and that have masses of up to ~$\sim 1$ TeV. 

Before discussing the implications of EWPO constraints for the LHC phenomenology of the xSM, we note that considerations of vacuum stability may eliminate some of the EWPO-compatible models with very light scalars (see, {\em e.g.}, Refs.~\cite{Davoudiasl:2004be,Casas:1996aq,Casas:1994qy,Hambye:1997ax,Hambye:1996wb,Altarelli:1994rb,Ford:1992mv,Sher:1988mj,Lindner:1988ww,Lindner:1985uk} . As indicated in Fig.~\ref{fig:sigreduce}, EWPO constraints eliminate many, but not all, of the models with a light $H_1$ and/or $H_2$. Although a detailed analysis of vacuum stability is beyond the scope of the present work, we emphasize that some of these remaining light-scalar models may be incompatible with vacuum stability. The lower bound of the Higgs mass depends on the top quark mass and especially on the cutoff scale $\Lambda$ of the theory.  In the case of the SM, for example, for the observed $m_t$ vacuum stability implies a lower bound on the Higgs boson mass that lies below the present LEP direct search bound for $\Lambda \gtrsim 10^6$ GeV, while the lower bound is $\sim 130$ GeV for $\Lambda\sim 10^{19}$ GeV \cite{Casas:1996aq,Casas:1994qy,Hambye:1997ax,Hambye:1996wb,Altarelli:1994rb,Ford:1992mv,Sher:1988mj,Lindner:1988ww,Lindner:1985uk}. The authors of Ref.~\cite{Davoudiasl:2004be} have studied a version of the xSM having $\mathbb{Z}_2$ symmetry and obtain a lower bound on the SM Higgs mass of $\sim 130$ GeV by requiring that the quartic couplings remain positive as the cutoff of the theory is increased to the Planck scale. An extensive study of both vacuum stability and perturbativity implications for the xSM will appear in a forthcoming publication \cite{us:2007xx}.

\section{Observation of a singlet mixed Higgs boson at the LHC}
\label{sect:statsig}

The SM Higgs boson is expected to be discovered at the LHC by combining a variety of channels \cite{ref:cmstdr,ref:atltdr}.  However, singlet mixing in the extended SM may spoil traditional signals due to a Higgs production strength that is weakened. The expected significances for 5$\sigma$ discovery of the SM Higgs boson at the LHC have been determined in simulations, with 100 fb$^{-1}$ at ATLAS and 30 fb$^{-1}$ at CMS \cite{ref:cmstdr,ref:atltdr} by considering a variety of observable modes.  ATLAS and CMS both utilize the gluon fusion production mode with the Higgs decays $H\to\gamma \gamma$, $H\to ZZ\to 4l$, $H\to WW\to l\nu l\nu$.  Additional modes in the analysis of ATLAS are $H\to ZZ \to l l \nu \nu$, $t \bar t H$ with $H\to b \bar b$ and the Higgstrahlung process $H W \to WWW\to l\nu l\nu l\nu$; the modes specific to the CMS analysis are the Weak Boson Fusion (WBF) processes $WW \to H$ with $H\to WW \to l \nu jj$, $H\to \tau \tau \to l+j$ and $H\to \gamma \gamma$.  The statistical significance of Higgs boson discovery in the xSM model is generated by scaling the significance of individual modes given by CMS by the fraction of a signal reduction $g_h^2/g_{h_{SM}}^2 \times \text{BF}(h\to X_{SM}) / \text{BF}(h_{SM}\to X_{SM})$ and summing the result in quadrature.  We emphasize the CMS search results here as we concentrate on distinguishing the SM Higgs sector and that augmented with a scalar singlet using early LHC data.  Therefore, we consider a Higgs boson as discoverable if its statistical significance is larger than $5\sigma$ for the 30 fb$^{-1}$ data at CMS.

\begin{figure}[t]
\begin{center}
\includegraphics[angle=-90,width=0.49\textwidth]{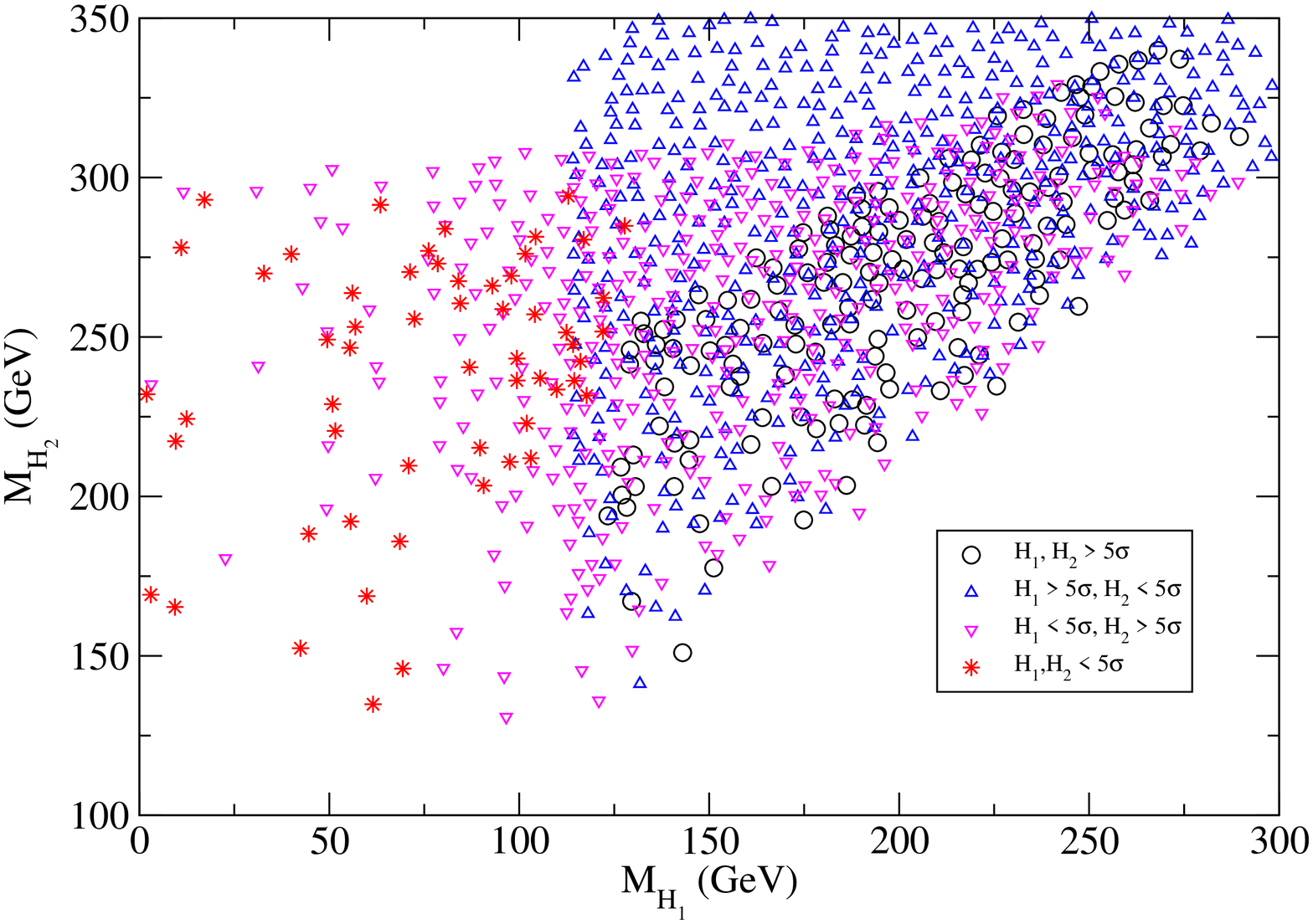}
\includegraphics[angle=-90,width=0.49\textwidth]{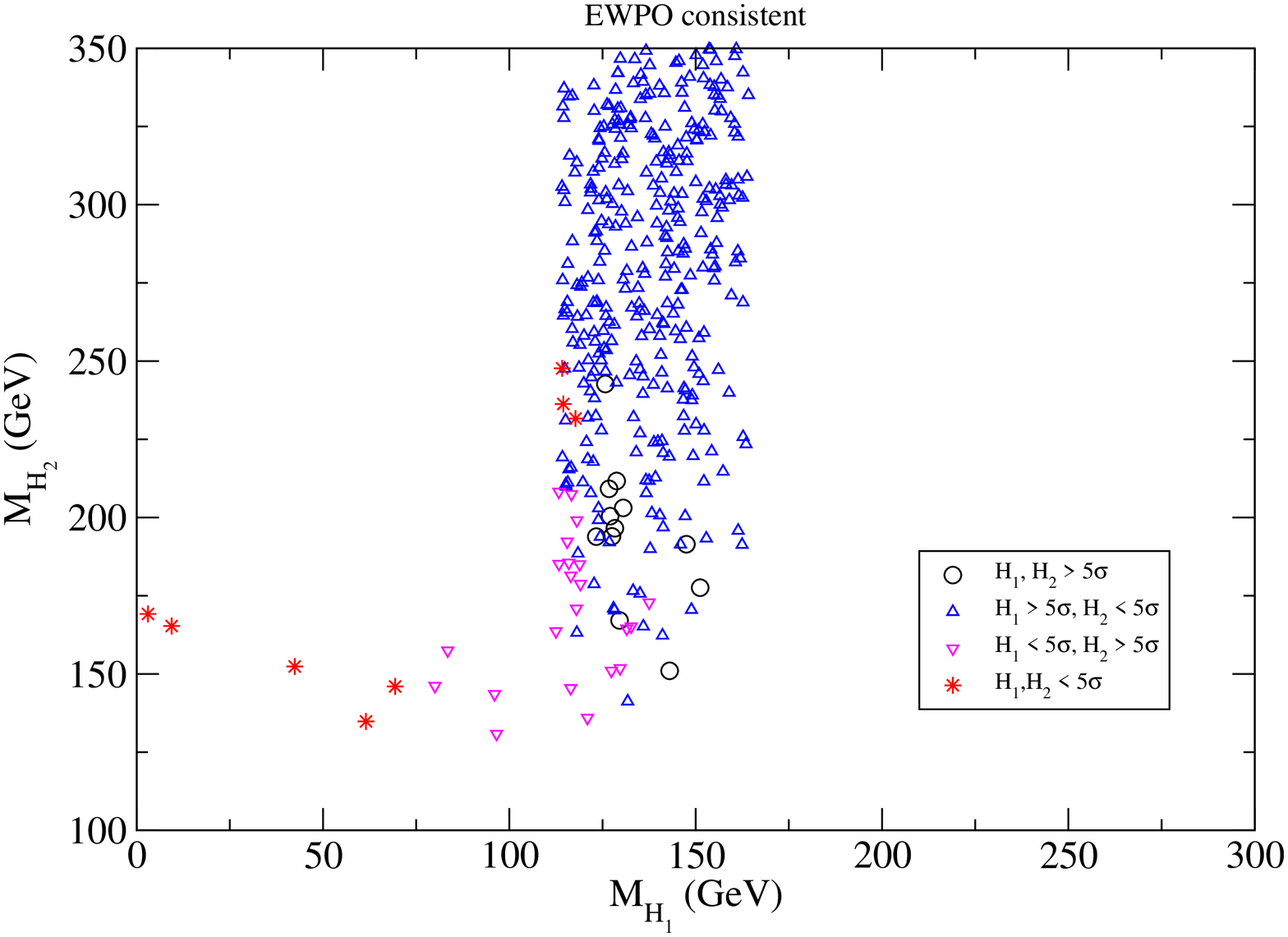}
\caption{Discovery potential (by traditional modes) of two Higgs bosons at CMS with 30 fb$^{-1}$ of data vs. the Higgs boson masses.  Shown are the points where both, one or no Higgs bosons are discoverable at the LHC.  Consistency with EWPO restricts the range of the lightest Higgs boson, making discovery of at least one Higgs boson likely.  Exceptions include the case of a light SM-like Higgs boson that dominantly decays to a light singlet-like Higgs.  The lightest Higgs can be very light due to the large singlet composition, making its coupling to SM fields weak.}
\label{fig:signif}
\end{center}
\end{figure}

In Fig. \ref{fig:signif}, we show the significance at CMS for the two Higgs states of this model. The left panel gives the results when only the LEP Higgs search constraints are imposed, while the right panel shows the corresponding discovery significance after EWPO constraints are applied.  In nearly all cases, at least one Higgs boson has a statistical significance that is above the $5\sigma$ level required for discovery.  The cases where both Higgs bosons are not discoverable are confined to the region where the statistical significance is not quoted for CMS (below $M_h=114$ GeV), or is under the $5\sigma$ significance required for discovery.  

\begin{figure}[t]
\begin{center}
\includegraphics[angle=-90,width=0.49\textwidth]{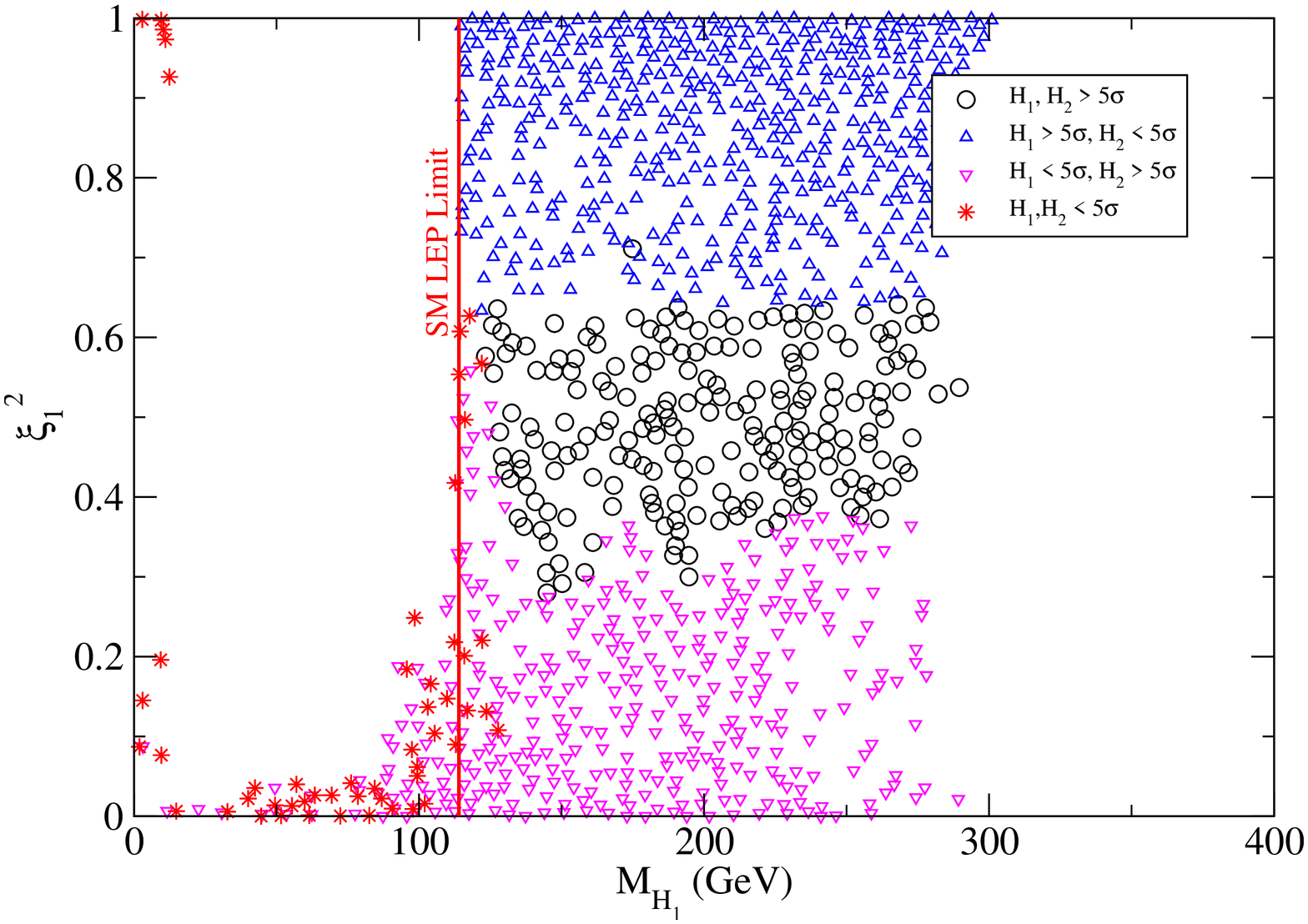}
\includegraphics[angle=-90,width=0.49\textwidth]{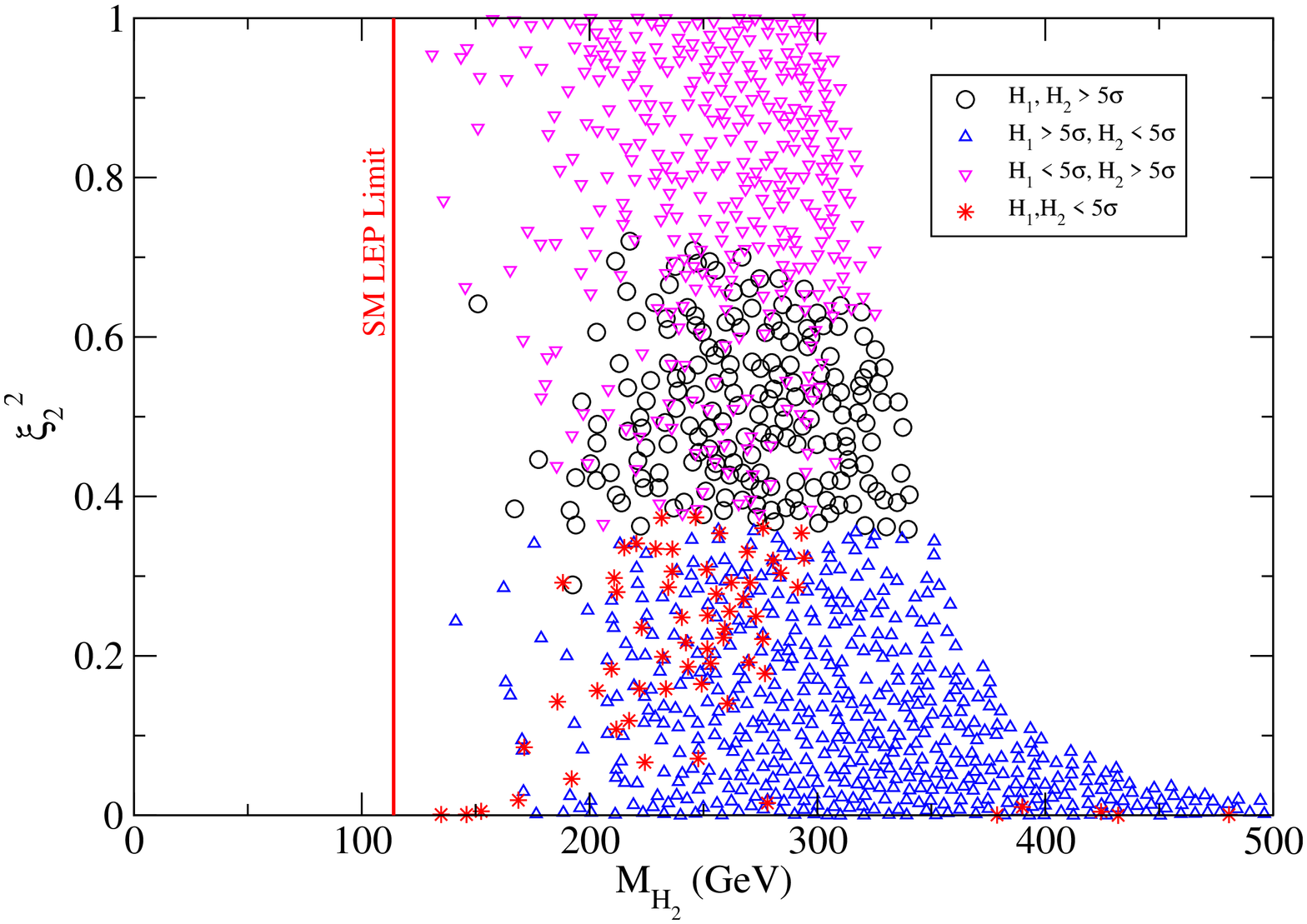}\\
\includegraphics[angle=-90,width=0.49\textwidth]{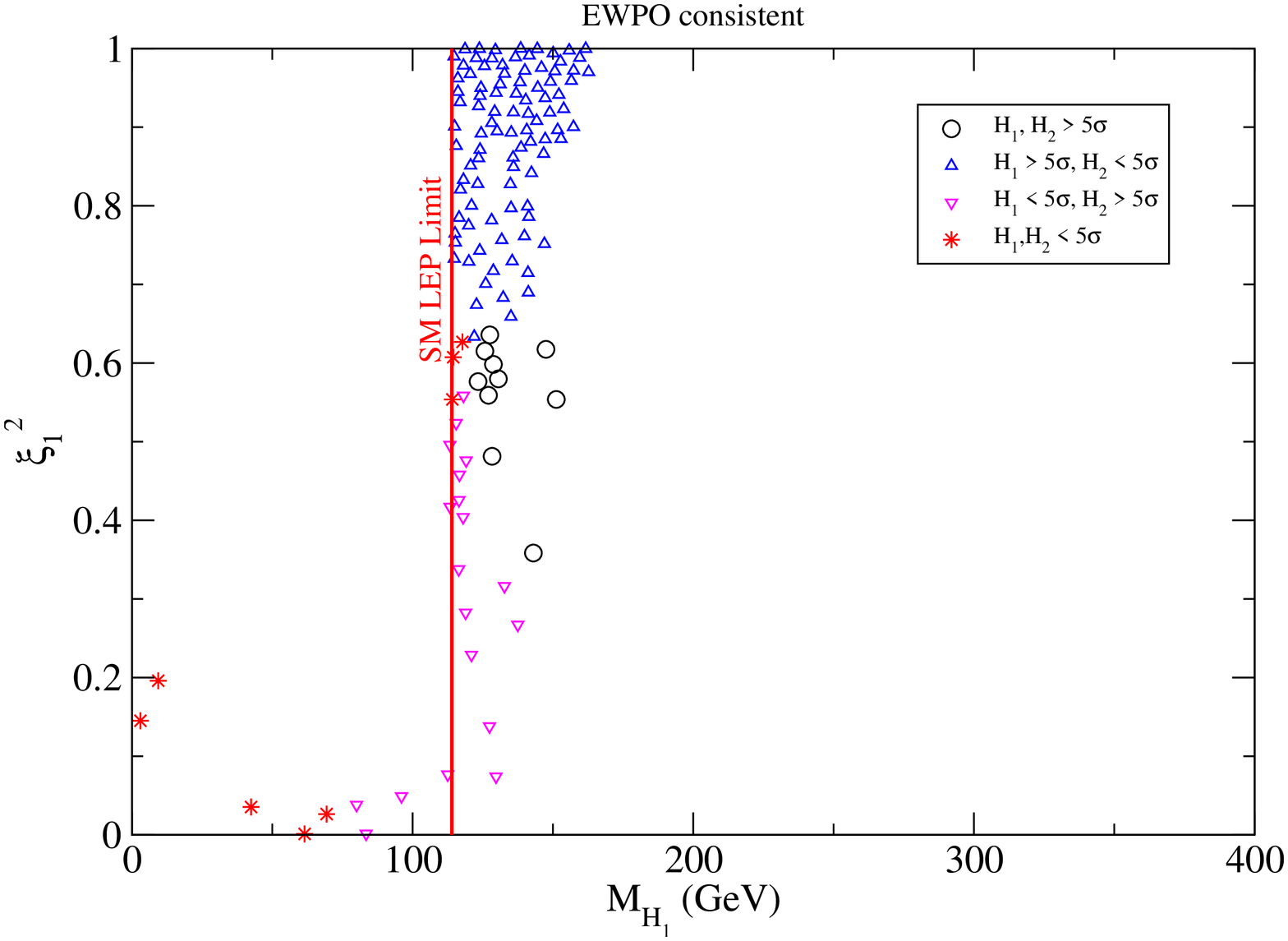}
\includegraphics[angle=-90,width=0.49\textwidth]{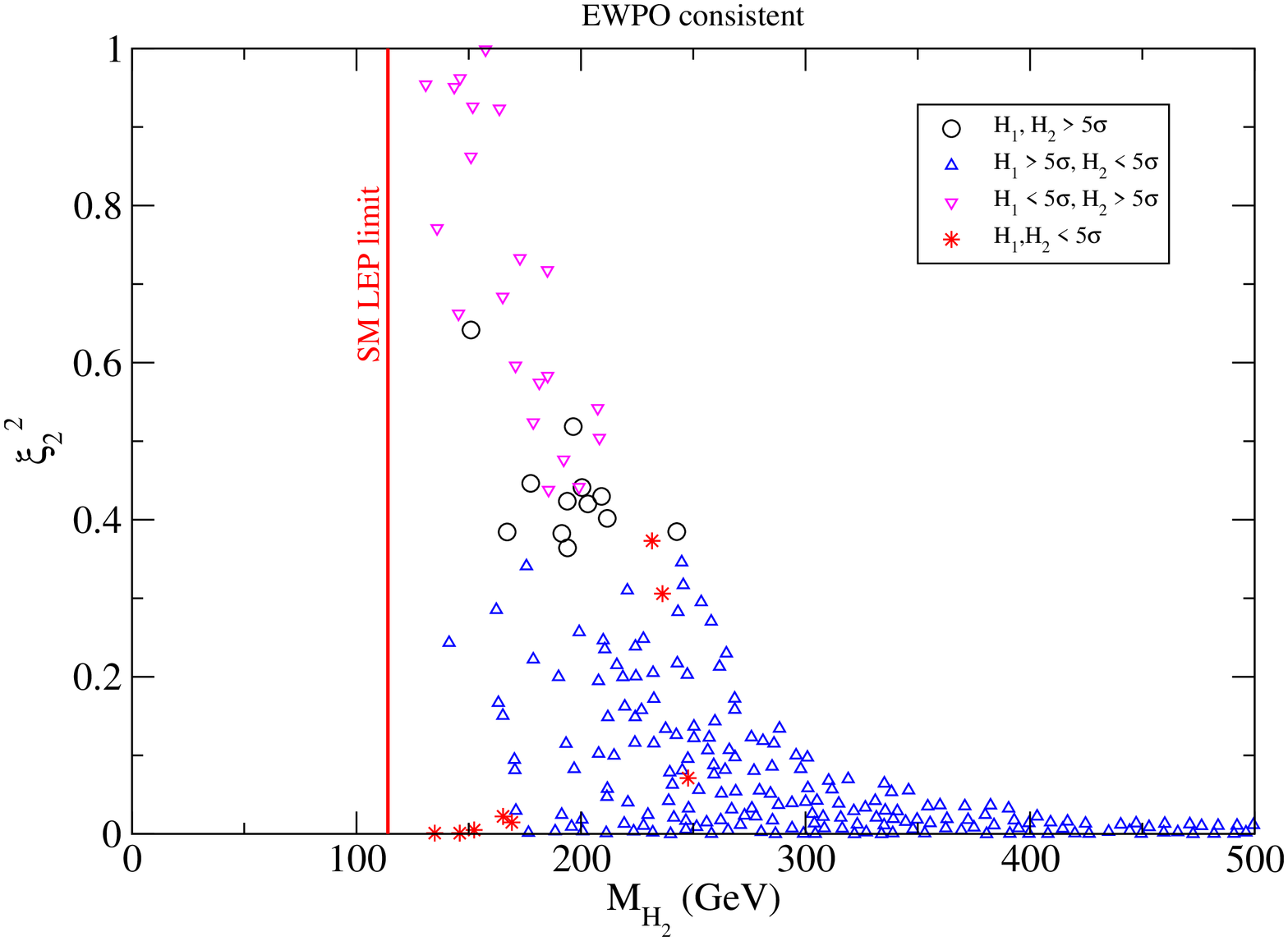}
\caption{Discovery potential of two Higgs bosons at CMS with 30 fb$^{-1}$ of data for masses and the signal reduction factor, $\xi^2$.  The left (right) panels show the mass and $\xi^2$ value for the light (heavy) Higgs boson while the bottom (top) panels do (do not) apply the constraints from EWPO.  Shown are the points where both, one or no Higgs bosons are discoverable at the LHC through the traditional modes.  Consistency with EWPO restricts the range of the lightest Higgs boson, making discovery of at least one Higgs boson likely.  Exceptions include the case of a light SM-like Higgs boson that dominantly decays to a light singlet-like Higgs.  }
\label{fig:coupstatsig}
\end{center}
\end{figure}

In the cases where the lightest Higgs boson is below the LEP SM Higgs mass limit, the second Higgs state can be discoverable as it has nearly full SM coupling strength.  This is evident in the points (pink triangles) of Fig. \ref{fig:coupstatsig}, where we show the discovery potential for the Higgs bosons of a given mass and signal reduction factor, $\xi^2$.  The left (right) panels show the mass and $\xi^2$ value for the light (heavy) Higgs boson while the bottom (top) panels do (do not) apply the constraints from EWPO.  Discovery of only one state with early data is possible if the mixing is not very strong, yielding a nearly decoupled singlet that does not produce a large signal.  As expected, to discover both Higgs states, the singlet-higgs mixing is required to be nearly maximal to allow a strong enough signal for both Higgs bosons.  However, cases where neither Higgs is discoverable with early data are possible where the SM-like Higgs boson dominantly decays to a light singlet-like Higgs. In this case, the SM-like Higgs has a reduced branching fraction to SM modes, reducing the effectiveness of the traditional search.  This is evident in the cluster of points where neither $H_1$ nor $H_2$ is discoverable (red stars) in the top right panel of Fig. \ref{fig:coupstatsig}.  These points have a large coupling factor that is given by (c.f. Eq. (\ref{eq:coup}))
\be
g_{H_2}^2 = 1-\xi_1^2 = 1-g_{H_1}^2,
\ee
meaning that the despite the large coupling strength of $H_2$, the decay to lighter scalars decreases the decays to traditional search modes.  However, with increased integrated luminosity beyond the 30 fb$^{-1}$ assumed, the prospects for discovery should improve.

The impact of EWPO on the discovery potential is quite pronounced. Models in which the lightest Higgs $H_1$ can be seen must have sufficiently large coupling ($\xi_1^2$) to SM modes and must also be lighter than $\sim 160$ GeV to be consistent with precision electroweak data as seen in Figs. \ref{fig:signif} and \ref{fig:coupstatsig}. If the EWPO constraints are not imposed (top panels), the existence of a significantly heavier $H_1$, that couples strongly enough to SM modes to be discovered is allowed. The EWPO exclude this possibility. Similarly, the set of models leading to an observable $H_2$ is considerably reduced by EWPO considerations, since any scalar must have a large enough $\xi_i^2$ to be seen in the conventional Higgs decay channels but must be light enough to satisfy the EWPO requirements. 

We note that many of EWPO-allowed models having a $\geq 5\sigma$ discovery significance using the conventional Higgs decay modes are those giving a  SM-like scalar that is the lightest of the two mass eigenstates with $M_{H_1}\lesssim 160$ GeV.  The presence of the second decoupled scalar does not alter the discovery potential of the light SM Higgs boson in these channels. However, the presence of the augmented scalar sector enhances the possibility for discovering an EWPO-compatible Higgs boson that is heavier than a pure SM Higgs. This possibility is most clearly illustrated by the bottom panels of Fig. \ref{fig:coupstatsig}. The bottom right panel contains numerous models that yield a $\geq 5\sigma$ discovery of the $H_2$ with masses ranging from $\sim 150$ to $\sim 220$ GeV. The value of $\xi^2_2$ decreases with $M_{H_2}$ as needed to satisfy the EWPO constraints. 

\begin{figure}[t]
\begin{center}
\includegraphics[angle=-90,width=0.69\textwidth]{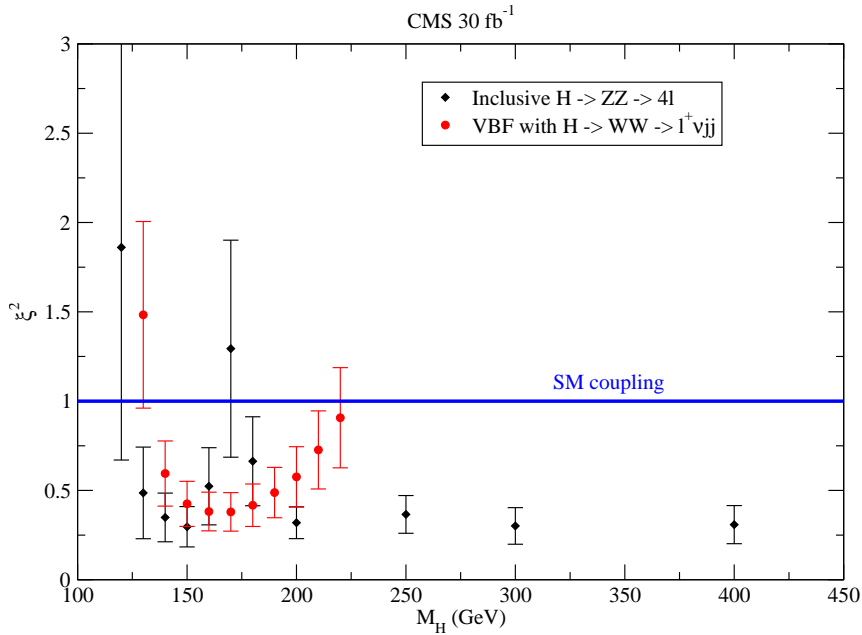}
\caption{Minimum value of $\xi^2$ that can be probed at CMS with 30 fb$^{-1}$ of data in discovering a Higgs boson with $5\sigma$ significance.  We consider the modes $H\to ZZ\to 4\ell$ and $H\to WW\to \ell \nu jj$.  Over most of the range $M_H \gtrsim 150$ GeV, the Higgs boson can be discovered and its mixing can be shown to significantly deviate from the SM value.  Points shown that are above the SM coupling line imply that an enhanced Higgs coupling is required to yield a $5\sigma$ signal.  The $ZZ$ mode is weakened at $m_H=170$ GeV since the Higgs decay becomes dominated by $H\to WW$.}
\label{fig:coupred}
\end{center}
\end{figure}

More generally,  if a light scalar is observed in these modes its couplings could be reduced from those of the SM Higgs, as suggested by the points in Fig. \ref{fig:sigreduce}.  It is interesting to ask how large this reduction can be while still yielding a $\geq 5\sigma$ significance in the conventional Higgs decay modes at the LHC.  To address this question, we show in Fig. \ref{fig:coupred} the minimum value of $\xi^2$ needed to yield a $5\sigma$ discovery in $H\to ZZ\to 4\ell$  and $H\to WW\to \ell^+\nu + jj$  with 30 fb$^{-1}$ at CMS.  The central value is obtained from the ratio of the number of events that require a $5\sigma$ signal in the xSM and the analogous number of events assuming a SM Higgs boson.  The $1\sigma$ error bars are due to the finite statistics and signal uncertainty of the reduced discovery signal~\footnote{Note that these results have been scaled from the cross section for the signal and background for the $WW$ mode and scaled for 10 fb$^{-1}$ of data quoted for the $ZZ$ mode.  Therefore, these results illustrate how well one can probe the $h-S$ mixing.}.  The results indicate that for scalars with mass $\gtrsim 150$ GeV, one could observe Higgs having $\xi^2$ as low as $\sim 0.4$ with $\geq 5\sigma$ significance and, for most of the allowed mass range, determine that its coupling is reduced from that of the SM Higgs.  Points shown that are above the SM coupling line imply that an enhanced Higgs coupling is required to yield a $5\sigma$ signal.  The $ZZ$ mode is weakened at $m_H=170$ GeV since the Higgs decay becomes dominated by $H\to WW$.  With more integrated luminosity, the measurement uncertainty decreases.  Individual Higgs boson couplings can be determined to an accuracy of order 10-30\% at ATLAS and CMS with a combined luminosity of 800 fb$^{-1}$\cite{Duhrssen:2004uu,Zeppenfeld:2000td}. Future studies at a $\sqrt{s}=500$ GeV Linear Collider could yield couplings with  2-5\%  precision  assuming couplings of SM strength~\cite{Abe:2001np}.  

An alternate signature of an augmented scalar sector would be the presence of the kinematically allowed decay $H_2\to H_1 H_1$. If $H_2$ is SM-like, then the presence of this decay mode would result in a reduced $H_2$ branching fraction to SM Higgs decay products \cite{O'Connell:2006wi}. From Eq.~(\ref{eqn:xi2def}) we observe that even if $S$ and $h$ do not mix appreciably and $g^2_{H_2}\approx1$, the parameter $\xi^2_2$ can still be much less than one due to the presence of the $H_2\to H_1 H_1$ channel. This possibility is particularly interesting from the standpoint of cosmology, since many EWPO-allowed models that also yield a strong first order EWPT are accompanied by reduced branching ratios of a SM-like Higgs to conventional SM modes \cite{Profumo:2007wc}. The results of Fig. \ref{fig:coupred} indicate that one could probe such models that lead to branching ratios as low as 40\% of the SM expectation. Models with larger reductions also lead to a strong first order EWPT, and they could be probed using greater integrated luminosity or with future Higgs decay studies at a Linear Collider.

Apart from observing a reduction of $\xi^2_2$ from unity, one could also search for exotic final states that result from the decays of the two $H_1$ bosons in the Higgs splitting channel. In this respect, the LHC phenomenology of the xSM can have features that resemble some singlet extended supersymmetric models.  In these extended models, the light CP-odd Higgs boson has been studied at length \cite{Barger:2006dh,Barger:2006sk,Dermisek:2006py,Dermisek:2006wr,Dermisek:2005gg,Dermisek:2005ar,Carena:2007xx,Cheung:2007sv}.  The presence of the Higgs splitting mode $H_2\to H_1 H_1$ may also lead to unusual final states, such as four $b$-jets, $b{\bar b}\tau^+\tau^-$, or $b{\bar b}\gamma\gamma$. The feasibility of observing these exotic states has been considered in the Higgsstrahlung production mode, $W/Z\to W/Z H_2\to W/Z+2H_1\to 4X+\ell\nu/\ell \ell$, where $4X$ denote the decay products of the two light scalars \cite{Cheung:2007sv,Carena:2007xx}.  Discovery for benchmark points were illustrated in the NMSSM with SM-like Higgs bosons of mass $\approx 110-120$ GeV decaying to two CP-odd Higgs bosons of mass $\approx 30-40$ GeV. The decays of the $H_1$  in the xSM are similar to the CP-odd Higgs decays of the NMSSM as they are dominated by fermionic decays and $\tan \beta$ does not appreciably change the ratio of  decays to $b\bar b$ and $\tau^+ \tau^-$.  Thus, the analogous scenario in the xSM is compatible with EWPO and may provide an observable signal of the Higgs splitting mode.  We note in passing that when the singlet-like $H_1$ is very light, the SM-like $H_2$ may have escaped detection at LEP if $M_{H_1} < 2 m_b$.  In this case, the search mode through the bottom channel is not accessible, making the $\tau^+ \tau^-$ channel dominant \cite{Dermisek:2006wr}. 

The prospects for identifying the complementary case, where $H_2$ is the singlet-like Higgs, by observing exotic final states are less definitive.  For the decay to two SM-like Higgs bosons to be kinematically accessible, the mass of $H_2$ must be $\gtrsim 220$ GeV since $H_1$ obeys the LEP limit on the SM Higgs mass.  In this case, the Higgstrahlung cross section is significantly decreased due to the $Z$-propagator suppression at high $\sqrt{\hat s}$.  In addition, the production cross section is reduced by the mixing factor $\sin^2\phi \le 0.5$.  A simple computation of the cross sections for producing exotic final states through the Higgs splitting channel of a singlet-like $H_2$ suggests the possibility of several hundred four $b$-jet events before cuts with 30 fb$^{-1}$ integrated luminosity. However, the practical feasibility of observing such events at the LHC requires a detailed analysis that goes beyond the scope of the present study, and  a Linear Collider may provide a more realistic prospect for observing these modes.

\section{The Singlet as Dark Matter}
\label{sect:dm}
If the scalar potential of Eq.~(\ref{eqn:hpot}) respects a 
$\mathbb Z_2$ symmetry, the singlet $S$ can be a viable dark matter candidate\footnote{It is difficult to obtain a strong first order electroweak phase transition in this case unless a large number of light singlets are present~\cite{Espinosa:2007qk}.}.  Assuming thermal DM production and standard cosmology, the parameters appearing in Eq.~(\ref{eqn:hpot}) that govern the mass of the $S$ and its coupling to the SM Higgs, $h$, should reproduce the observed DM relic density. Here, we determine the regions of the xSM parameter space that satisfy these considerations, compute the corresponding nuclear recoil direct detection cross sections, and analyze the implications for Higgs boson searches at the LHC.

\subsection{Relic density}
\label{sect:scatt}

The relic density of dark matter has been determined by the WMAP 3-year CMB data and the spatial distribution of galaxies to be $\Omega_{\rm DM} h^2=0.111\pm 0.006$, where $h=0.74\pm 0.03$ is the Hubble constant~\cite{Spergel:2006hy,Yao:2006px} .  The relic density of dark matter relative to the closure density is very roughly given (for thermal production) by the total annihilation cross-section by $\Omega_{DM}h^2 \simeq 0.1\text{ pb}/\langle\sigma_{\rm ann} v\rangle$, where $v$ is the relative velocity~\cite{Yao:2006px}.  To precisely calculate the singlet relic density, we integrate the Boltzman equation according to Refs. \cite{McDonald:1993ex,Kolb:1990vq,Gondolo:2004sc}.  

\begin{figure}[htbp]
\begin{center}
\includegraphics[angle=0,width=0.16\textwidth]{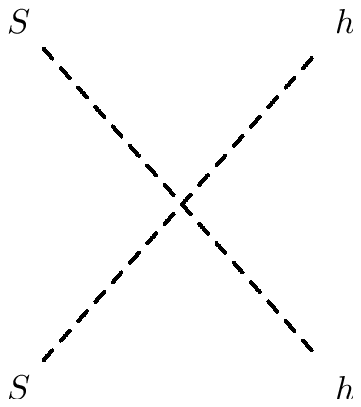}~~~
\includegraphics[angle=0,width=0.19\textwidth]{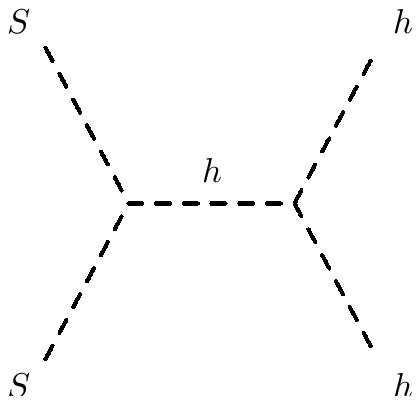}~~~
\includegraphics[angle=0,width=0.19\textwidth]{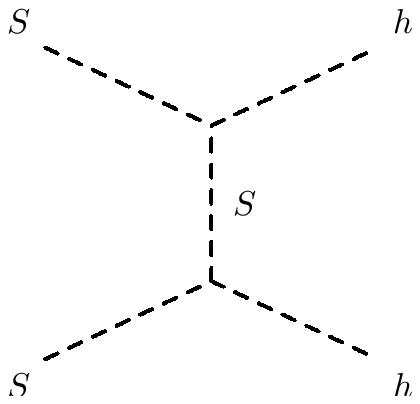}~~~
\includegraphics[angle=0,width=0.19\textwidth]{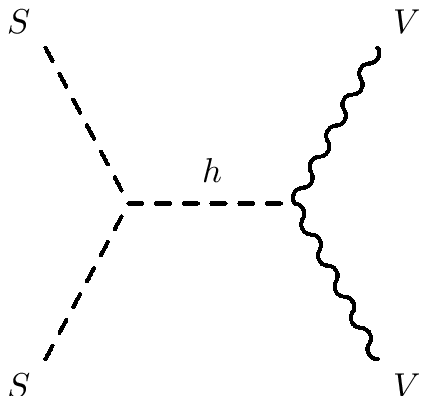}~~~
\includegraphics[angle=0,width=0.19\textwidth]{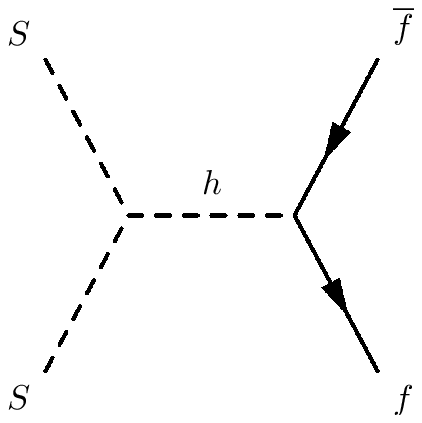}
\caption{Annihilation processes that contribute to the thermally averaged cross section.  All processes are mediated via the Higgs boson.}
\label{fig:annfd}
\end{center}
\end{figure}

The thermally averaged annihilation cross section is determined from the contributions of the processes shown in Fig. \ref{fig:annfd}.  Since all of the processes involve the SM Higgs boson, $h$, the key parameters in obtaining the observed relic density are $\delta_{2}$ and $\lambda$.  The $s$-channel Higgs couples to the usual SM final states. The $SS\to h\to hh$ diagram is mediated by the Higgs self coupling, although this diagram is expected to be suppressed since the intermediate $s$-channel Higgs boson is far off-shell.  We calculate the relic density of singlet dark matter in this model for the parameter ranges given in Eq.~(\ref{eq:ranges}).

\begin{figure}[htbp]
\begin{center}
\includegraphics[angle=-90,width=0.49\textwidth]{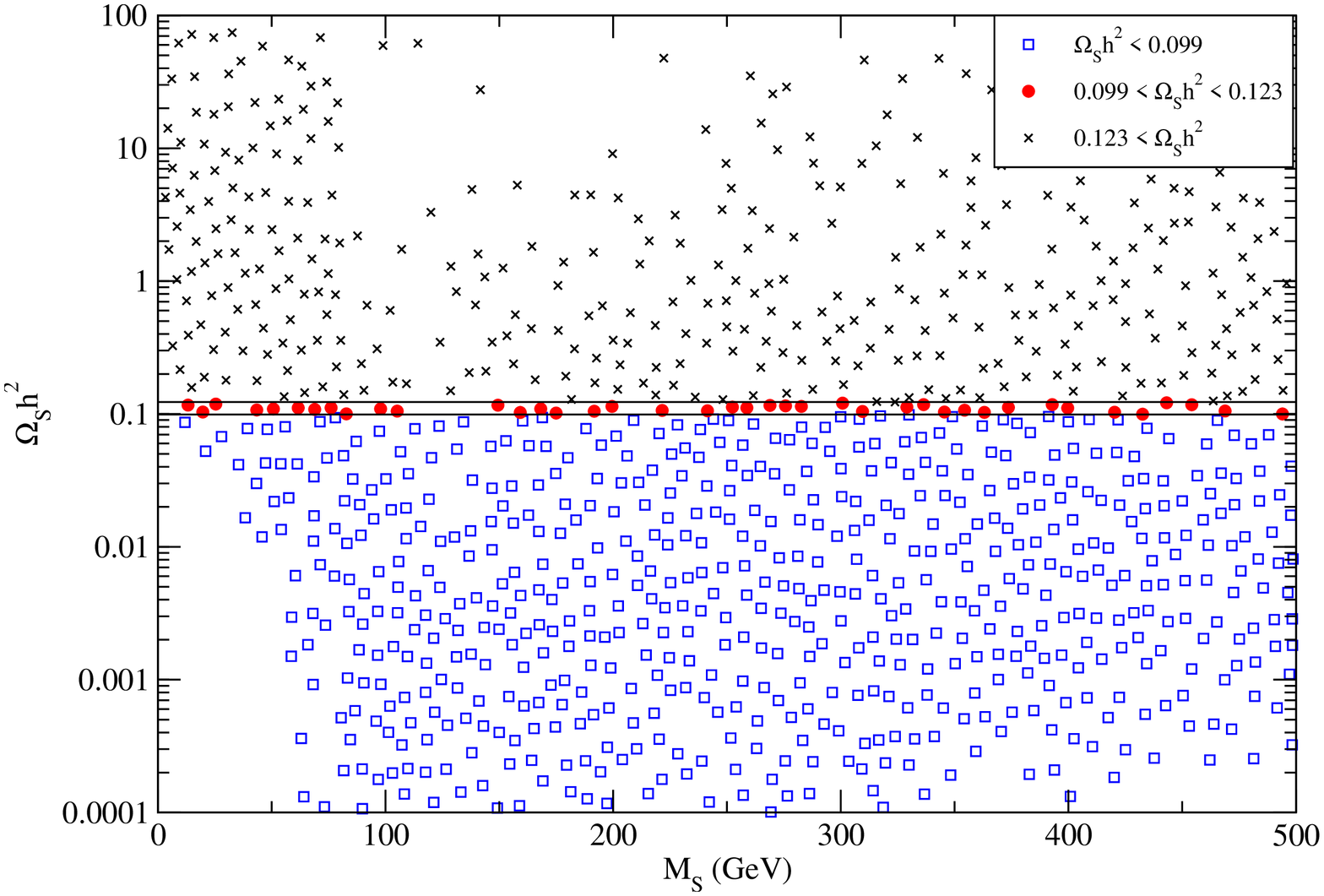}
\includegraphics[angle=-90,width=0.49\textwidth]{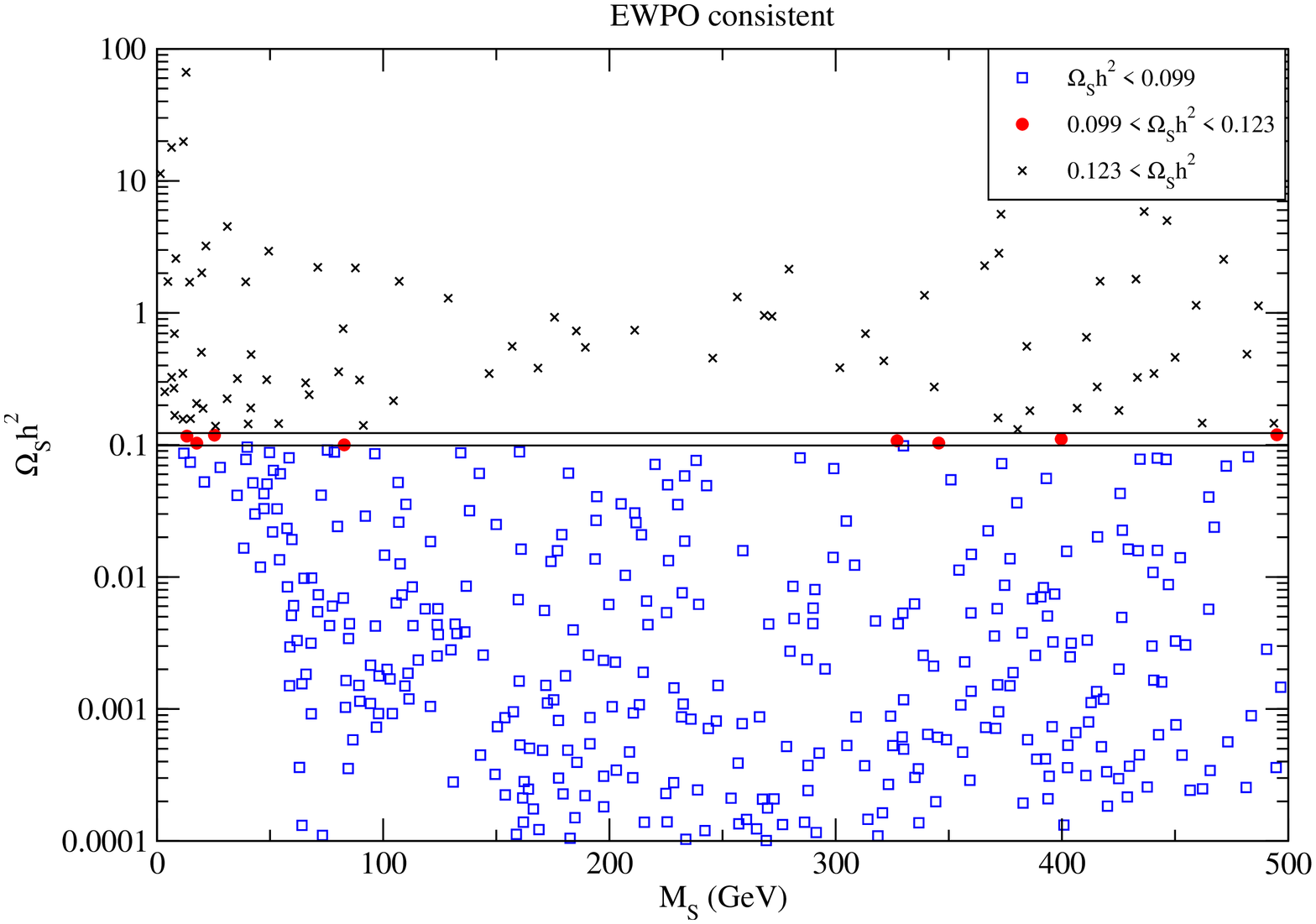}
\caption{Relic density of singlet dark matter versus the singlet mass with (right) and without (left) EWPO constraints on the Higgs boson mass applied.}
\label{fig:rdms}
\end{center}
\end{figure}

The relic density of the singlet DM is shown in Fig. \ref{fig:rdms} versus the singlet mass with (right) and without (left) EWPO consistency, which for the SM higgs boson implies $M_h < 150$ GeV~\footnote{In the LEP Electroweak Working Group fit that does not include low-energy EWPO, the upper limits from the EWPO fits are relaxed to $M_h < 182$ GeV when one includes the direct lower limit constraints \cite{LEPEWWG:2007}. }.  Imposing this bound can severely restrict the space of models.  Fewer points are shown in the observed range and appear to be non-uniform after imposing the EWPO constraint on the Higgs mass, but these are due to the small window of $\Omega_{DM}h^2$ and the limitations of the scan.  The observed region of the relic density allows a wide range of singlet masses. As the singlet mass is decreased below $M_h/2$, the annihilation cross section is suppressed since the singlets annihilate through an off-shell Higgs boson.  In this case, the minimum relic density increases sharply, as seen in Fig. \ref{fig:rdms} for $M_S \lesssim 50$ GeV.

\begin{figure}[t]
\begin{center}
\includegraphics[angle=-90,width=0.49\textwidth]{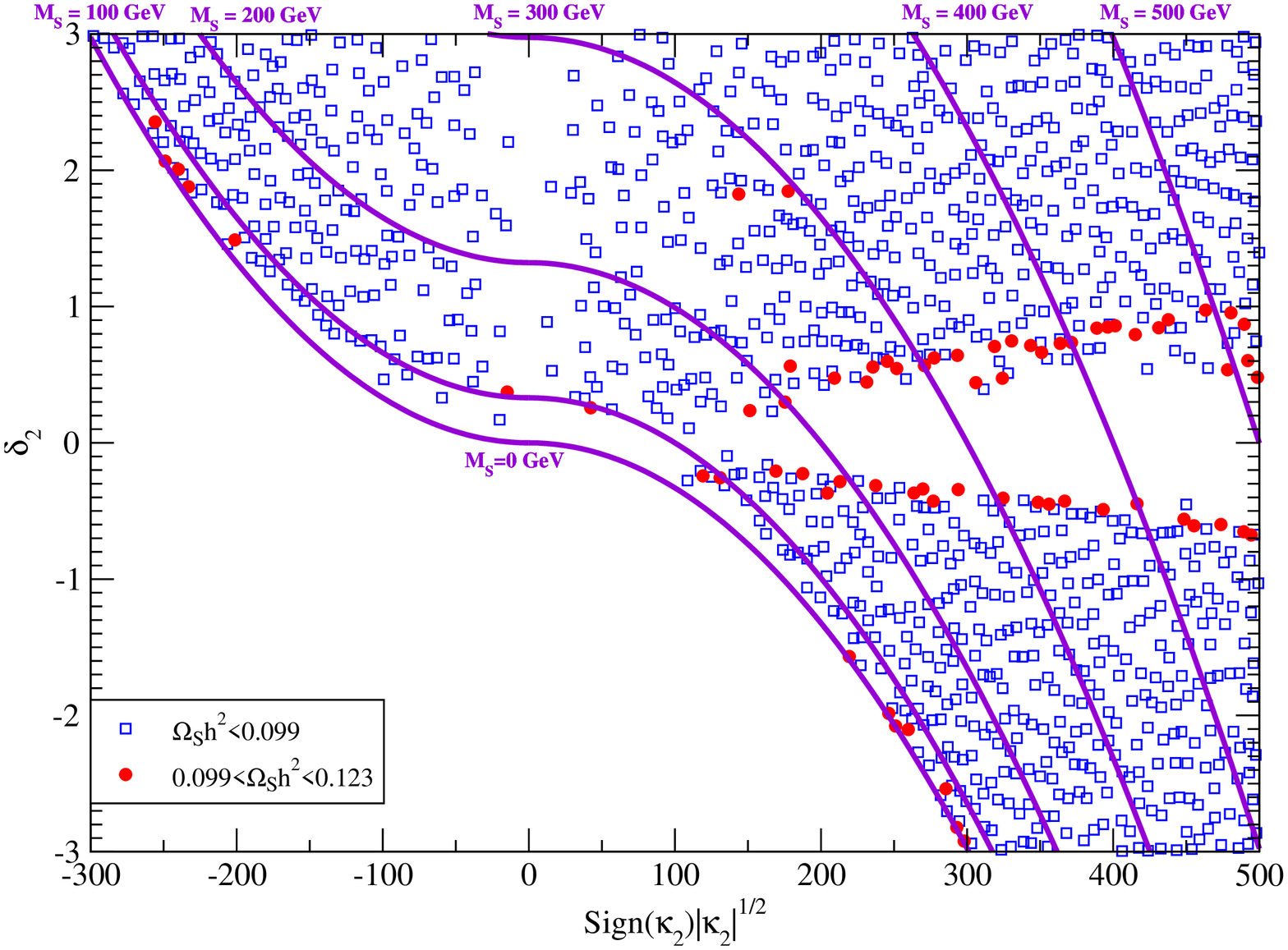}
\includegraphics[angle=-90,width=0.49\textwidth]{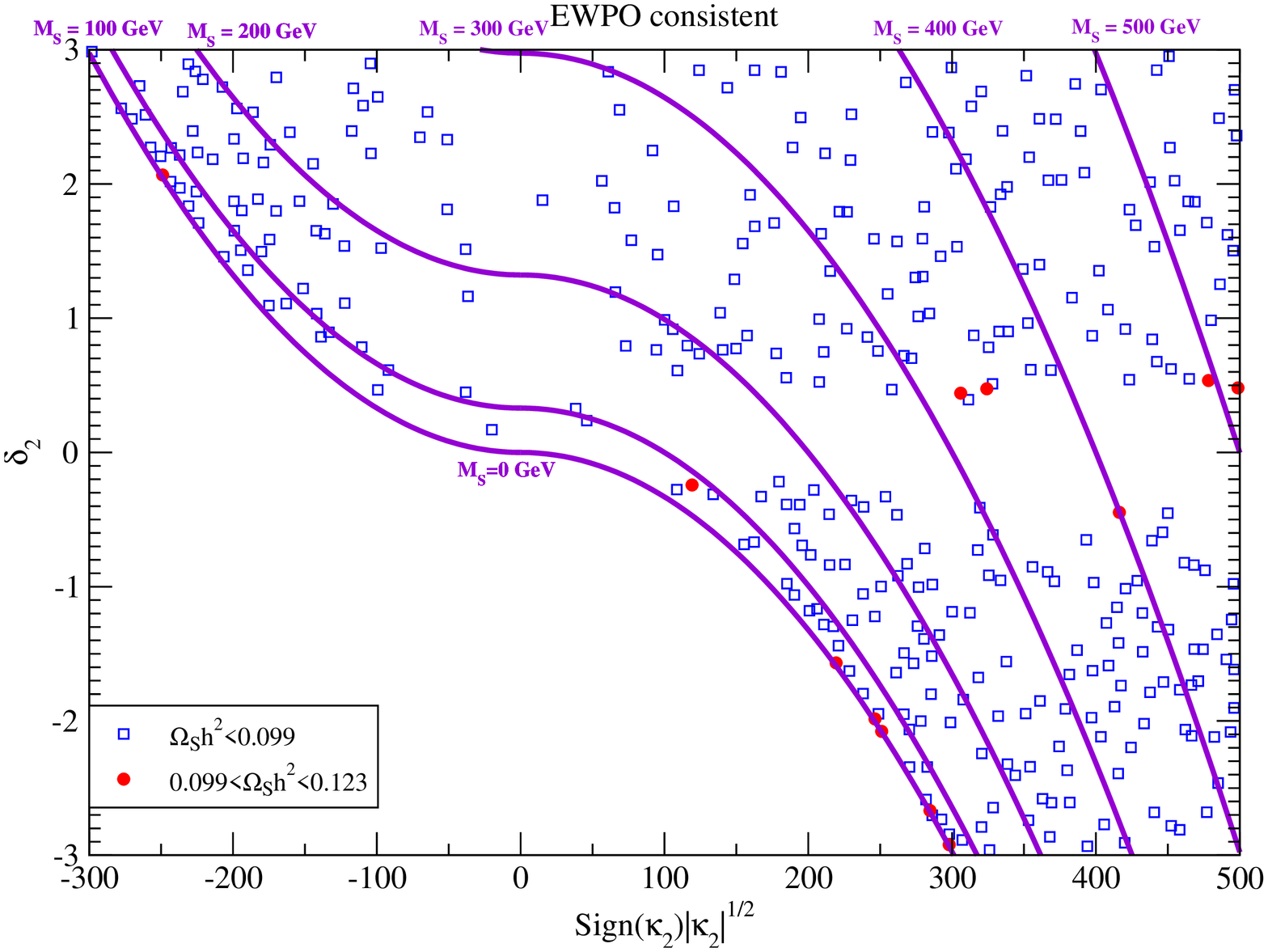}
\caption{Predicted relic density values in the plane of $\sqrt{\kappa_{2}}$ and $\delta_2$ with (right) and without (left) EWPO constraints on the Higgs boson mass applied.  Since the DM mass scales with $\sqrt{\kappa_2}$ for large $\kappa_2 \gtrsim \delta_2 v^2$, we show the parameter $\sqrt{\kappa_2}$ to illustrate the dependence of the relic density on the singlet mass.  Singlet DM masses are given by contours.  The open region with small $|\delta_2|$ and large $\kappa_2$ correspond to models (not shown explicitly) yielding an overdensity of relic DM. }
\label{fig:kap2del2}
\end{center}
\end{figure}

The relic density is largely affected by the Higgs-singlet coupling, $\delta_2$.  The singlet mass is determined uniquely by the parameters $\delta_{2}$ and $\kappa_{2}$.  Therefore, the region of parameter space consistent with the observed relic density values is restricted.  This small region does span a large range of singlet masses as shown in Fig. \ref{fig:rdms}.  We show in Fig. \ref{fig:kap2del2} the ranges of these parameter values that are consistent with the observed relic density with contours tracing out the values of the singlet mass with and without EWPO constraints.  It is evident that the limit of small $|\delta_2|$ leads to an excess of relic singlet DM since the annihilation rate is too small.  Conversely, large values of $|\delta_2|$ lead to an under-density assuming thermal production~\footnote{The relic density can be substantially enhanced if a non-thermal mechanism is present, or if there exists some other non-standard cosmological scenario \cite{Salati:2002md,Profumo:2003hq,Rosati:2003yw,Chung:2007vz}.}.   As the mass of the singlet increases, this coupling must be increased to maintain a relatively fixed annihilation cross section corresponding to a fixed relic density.  This trend is evident from the red circles in Fig. \ref{fig:kap2del2}.  The distinction between the different regions of relic density is blurred due to the variation of the Higgs mass through the scanned coupling $\lambda$.

\subsection{Direct Detection of Singlet Scalar Dark Matter: Elastic Scattering}
\label{sect:scatt}

\begin{figure}[b]
\begin{center}
\includegraphics[width=0.22\textwidth]{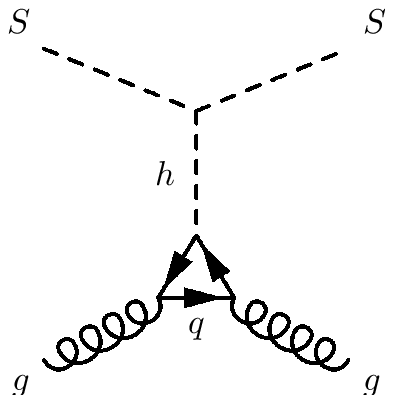}~~~~~~~~~
\includegraphics[width=0.19\textwidth]{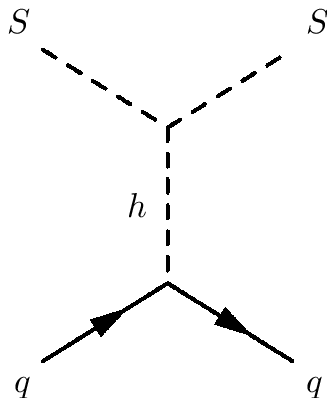}
\caption{Feynman diagrams for elastic scattering of the singlet DM particle off a proton.  The Higgs boson mediates the interaction.}
\label{fig:ddfd}
\end{center}
\end{figure}

Direct detection of dark matter can help establish the connection between dark matter and the model beyond the SM that is responsible for its existence.  The most promising prospect of detecting relic singlet scalar dark matter is via measuring its spin-independent interaction with nucleons.  Since the $S$ only interacts with matter via $t$-channel Higgs exchange (see Fig.~\ref{fig:ddfd} for the scattering processes), a sizable singlet-Higgs interaction is necessary to yield a positive signal.  Therefore, there is a very close relationship between Higgs physics and the direct detection of singlet dark matter.  Due to this close relationship, once the Higgs boson is found at the LHC, it may be possible to correlate the Higgs signal with the expected scattering rates of singlet DM.  Many studies have been made examining the relationship between collider and direct detection experiments \cite{Baltz:2006fm,Carena:2006nv} in supersymmetric models.

A number of experimental groups are carrying out direct detection experiments for spin-independent and spin-dependent scattering using both cryogenic and non-cryogenic methods \cite{Seidel:2005kx,Morales:2005ky}.  Limits on the spin-independent scattering cross sections have recently been reported by  CDMS \cite{Akerib:2006rr,Akerib:2005za,Akerib:2006ri,Akerib:2005kh}, EDELWEISS \cite{Sanglard:2005we}, WARP \cite{Benetti:2007cd} and Xenon10 \cite{Angle:2007uj}.  The latter uses a 15 kg liquid Xenon scintillator, and places a limit on the scattering cross sections on the order of $10^{-8}$ pb.  Future experiments like CDMS(2007) and SuperCDMS \cite{Akerib:2006rr} expect lower sensitivities to spin-independent interactions.  A summary of some current and future experimental sensitivities is given in Table \ref{tbl:siexp}.

\begin{table}[t]
\begin{center}
\caption{Spin-independent elastic scattering cross sections reach of various past and future experiments.  The maximal reach is for a light DM particle, typically $M_{DM}\approx 50$ GeV.}
\begin{tabular}{|c|c|}
\hline
Experiment&$\sigma_{DM-p}^{SI}$\\
\hline
CDMS \cite{Akerib:2005za}&$1.6\times10^{-7}$ pb\\
XENON10 \cite{Angle:2007uj}&$4.5\times10^{-8}$ pb\\
\hline
CDMS(2007) \cite{Ni:2006fh}&$1\times10^{-8}$ pb\\
WARP (140 kg) \cite{WARP} & $3\times10^{-8}$ pb \\
SuperCDMS (Phase A) \cite{Akerib:2006rr}&$1\times10^{-9}$ pb\\
WARP (1 ton) \cite{Aprile:2002ef}& $2\times10^{-10}$ pb\\
\hline
\end{tabular}
\end{center}
\label{tbl:siexp}
\end{table}%

In this xSM model the spin-dependent (SD) scattering cross section vanishes since there are no vector-like interactions that connect the singlet to matter.  Generally, for scalar DM, one cannot construct a SD coupling to SM fields.  In this case, if a positive SD signal is found by future experiments such as COUPP  \cite{Bolte:2006pf} or PICASSO \cite{Aubin:2006rc}, the xSM would be immediately ruled out as a viable DM scenario.  

In order to determine the sensitivity of present and future SI direct detection experiments to scalar singlet DM, we compute the corresponding SI scattering cross-section of a scalar dark matter particle off a nucleon:
\bea
\label{eq:scattrate}
\sigma^{SI}_{DM}={1\over 8 \pi (m_N+m_{DM})^2}{\delta_2^2 m_N^4\over M_h^4}
\left|\sum_{q=u,d,s} y_q f^p_{Tq}+\sum_{q=c,b,t} {2\over 27} y_q f^p_{TG}\right|^2
\eea
where the hadronic matrix elements, $f^p_{Tq}$ and $f^p_{TG}$, are given in Ref. \cite{Ellis:2000ds}.  Here, $y_q$ are the quark Yukawa couplings and $m_N$ is the nucleon mass.  The dominant contribution for SI scattering is due to $t$-channel Higgs exchange.  Since the cross-sections for scattering off protons and neutrons are very similar in size we calculate the scattering from protons.  We note that the uncertainty in the SI scattering cross-section is large, of order 60\%, due to the uncertainties in the hadronic matrix elements \cite{Ellis:2000ds}. 

\begin{figure}[htbp]
\begin{center}
\includegraphics[angle=-90,width=0.49\textwidth]{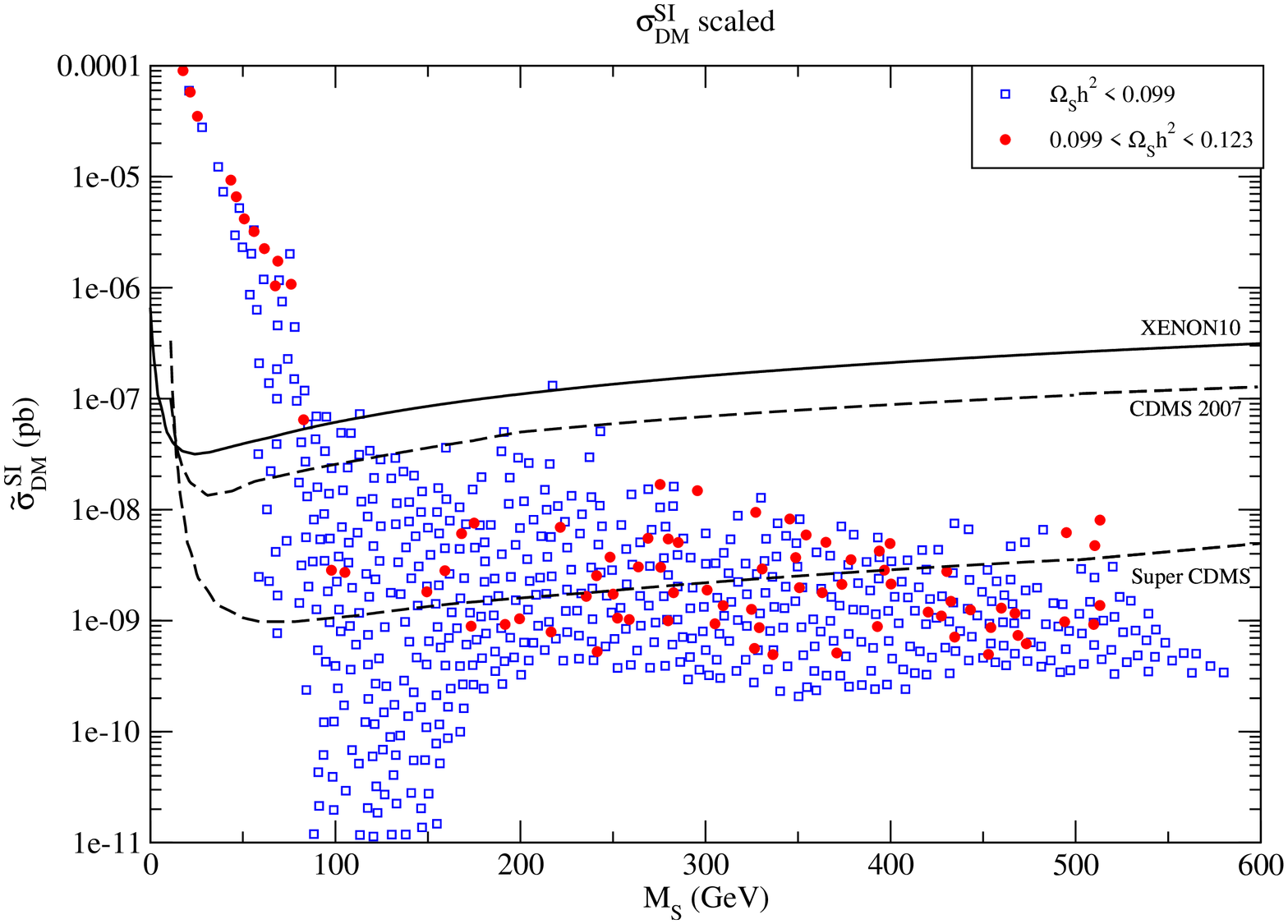}
\includegraphics[angle=-90,width=0.49\textwidth]{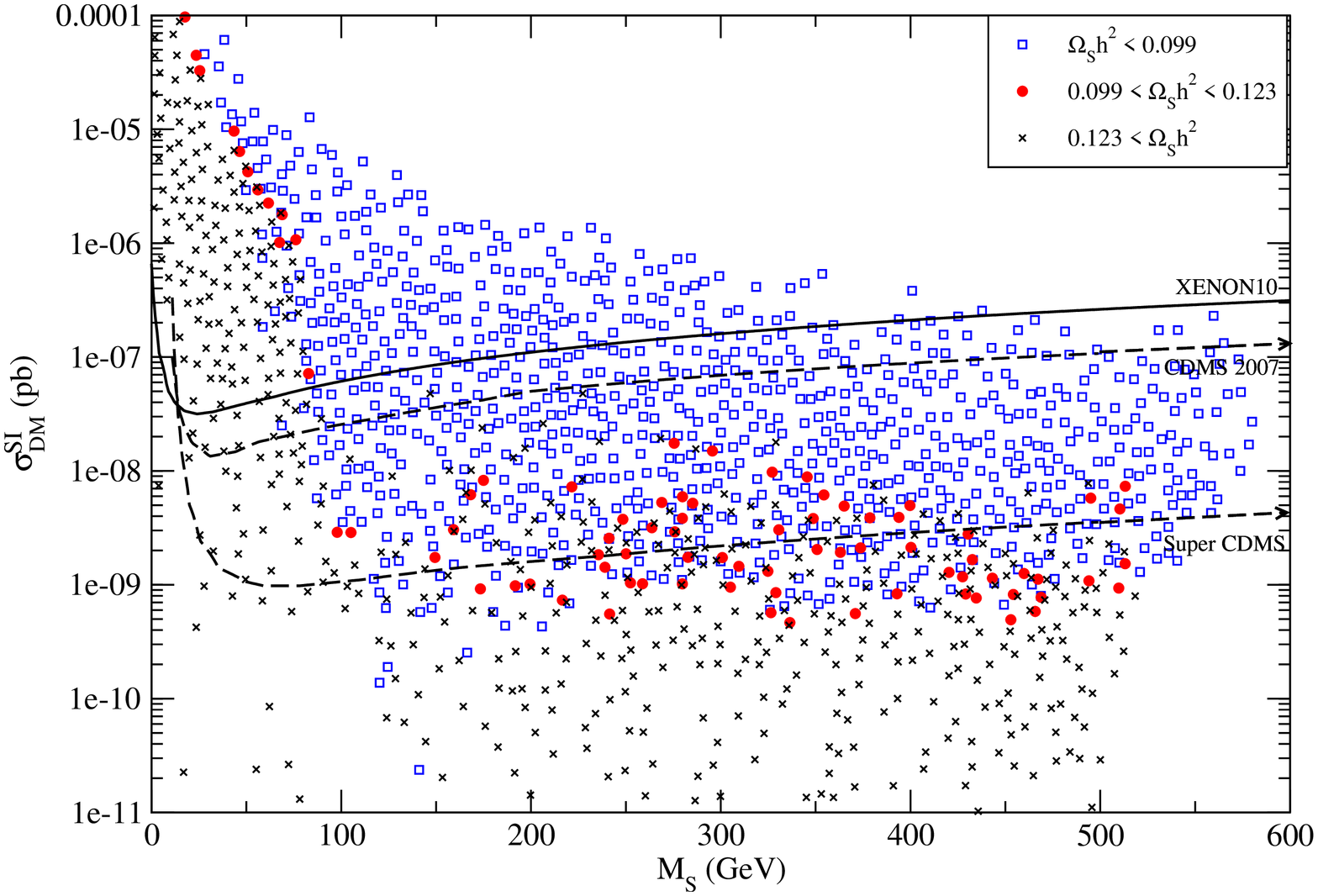}
\includegraphics[angle=-90,width=0.49\textwidth]{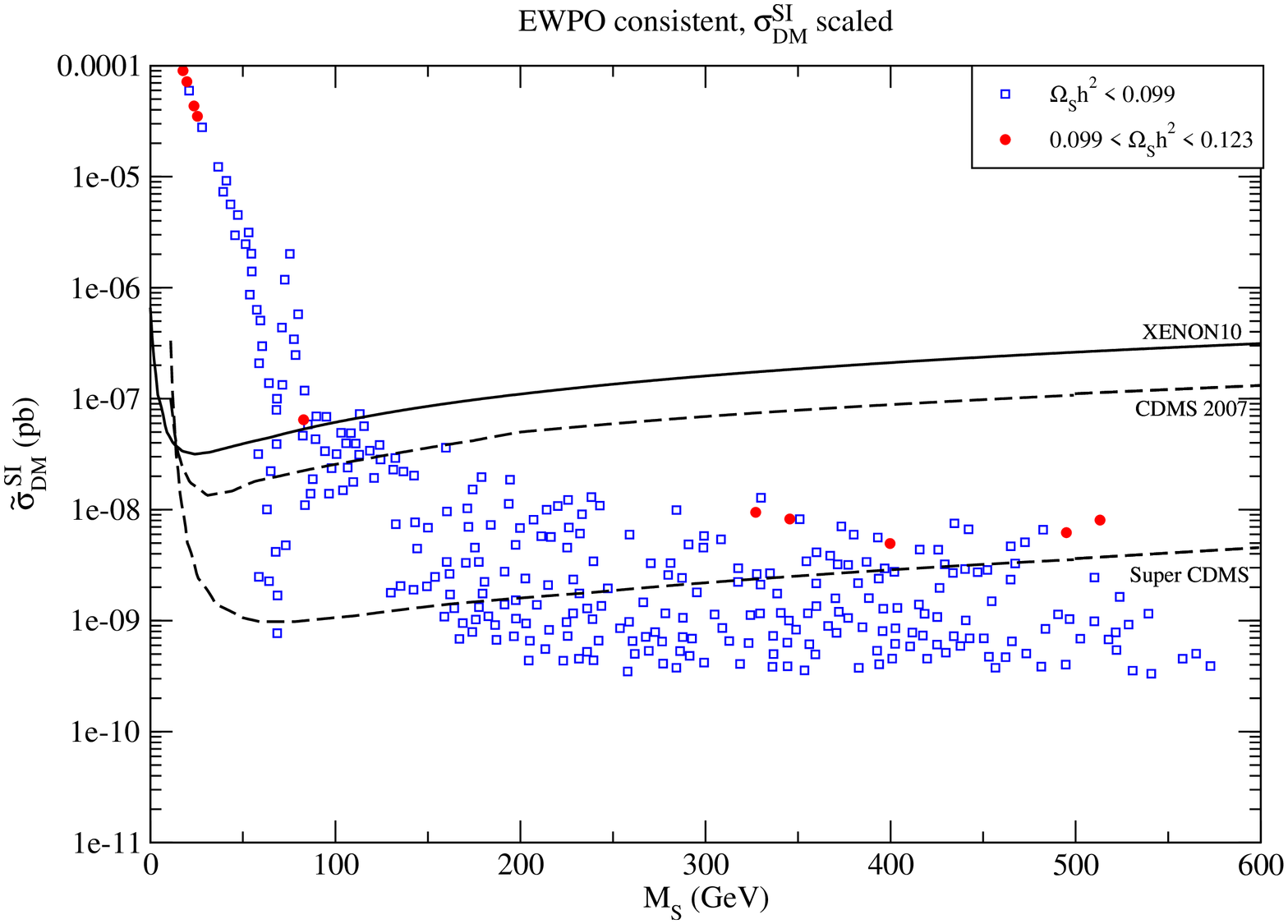}
\includegraphics[angle=-90,width=0.49\textwidth]{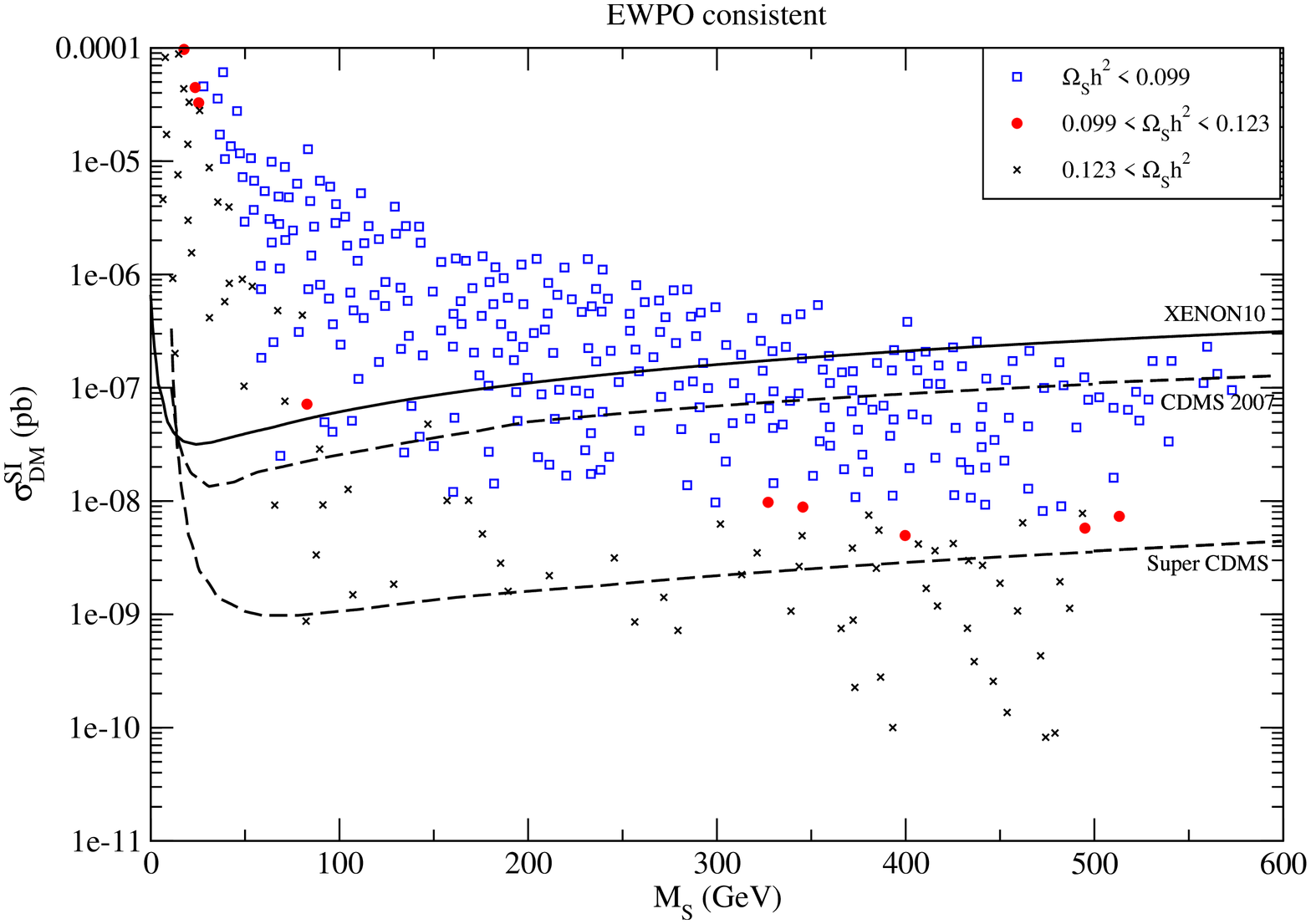}
\caption{Spin-independent cross section scaled (left) and not scaled (right) with the local density of dark matter for various DM masses.  The current best limit on the scattering cross section is from Xenon10 (solid).  It is expected that SuperCDMS will cover most of the scanned parameter space that is consistent with EWPO, with the exception that the singlets annihilate via the Higgs boson $s$-channel process (shown as points below SuperCDMS expected limit near $M_S \approx 100-150$ GeV).}
\label{fig:cross}
\end{center}
\end{figure}

The observed and expected limits given by various direct detection collaborations usually assume a local density of DM to be $0.3\text{ GeV/cm}^3$.  Therefore, we scale the scattering cross section by the ratio 
\be
\widetilde \sigma^{SI}_{DM} = \sigma^{SI}_{DM}\times{\Omega_{S}h^2 \over \Omega_{DM-WMAP} h^2}
\ee
to account for cases when there is predicted to be a deficit of dark matter in the universe in the singlet model~\footnote{The total amount of DM can still be consistent with the observed values by contributions from other sources, such as the axion.}.  The predicted spin-independent scattering cross section, scaled to the local DM density from a proton target, is shown in the left panels of Fig. \ref{fig:cross}.  Also shown are the existing cross section limits and expected sensitivities from a number of present and future direct detection experiments.  

In the event that a DM signal is not observed in the $10^{-9}-10^{-7}$ pb range that is expected to be probed by future experiments, it is still possible the singlet model is consistent with the observed relic density.  For example, singlet annihilation through the $s$-channel Higgs boson resonance implies the singlet-Higgs coupling must be substantially reduced to counter the resonant enhancement of the annihilation cross section in the relic density.  This feature is illustrated  by the points in the upper left panel of Fig.~\ref{fig:cross} having $M_S\approx 100-150$ GeV with scattering cross section below $10^{-9}$ pb. For these models, $m_H\approx 2M_S$, and the resonant annihilation cross section can be large enough to yield $\Omega_s h^2\leq 0.123$ even for small $\delta_2$. The resulting scaled spin-independent scattering cross sections are significantly suppressed. Away from the resonant condition, the same value of $\delta_2$ would lead to an over density due to the small cross section. This effect does not arise for $M_S\gtrsim 150$ GeV because our scan does not yield values of $m_h \gtrsim 300$ GeV. Excluding this region, many of the models provided by our scan could be probed by the SuperCDMS experiment. 

If one applies constraints from EWPO, the allowed range of scattering cross section that is below the present bounds shrinks since the favorably light Higgs boson generically has a large scattering rate (e.g., see Eq (\ref{eq:scattrate})).  In particular, this constraint removes a majority of the points where the $s$-channel Higgs pole necessitates a smaller $h$-$S$-$S$ coupling to achieve consistency with the relic abundance.  Overall, requiring EWPO consistency, the current phase of direct detection experiments should be able to cover a sizable part of the parameter space, if not probe the model completely.  

If, however, the relic density of DM is altered by either some non-thermal process or other non-standard cosmological scenario, one may assume that the local density of DM is fully accounted by the scalar singlet~\footnote{Typically, in these scenarios, the relic density is enhanced.  However, it may be suppressed in the case of a low reheating scenario~\cite{Giudice:2000ex}.}.  To analyze the direct detection in this case, we do not scale the scattering cross section (shown in the right panels of Fig. \ref{fig:cross}) since the scalar singlet  already saturates the observed relic density by these non-standard scenarios.  The points that evade the present limits from XENON10 that require an enhancement of relic DM may be eliminated if there is not a positive signal from future experiments.  However, in models giving a suppressed annihilation cross section, future experiments may not yield a positive signal due to a very small $h-S-S$ coupling.

\subsection{Impact on LHC searches for Higgs bosons}
\label{sect:higgsdecay}

\begin{figure}[t]
\begin{center}
\includegraphics[angle=-90,width=0.49\textwidth]{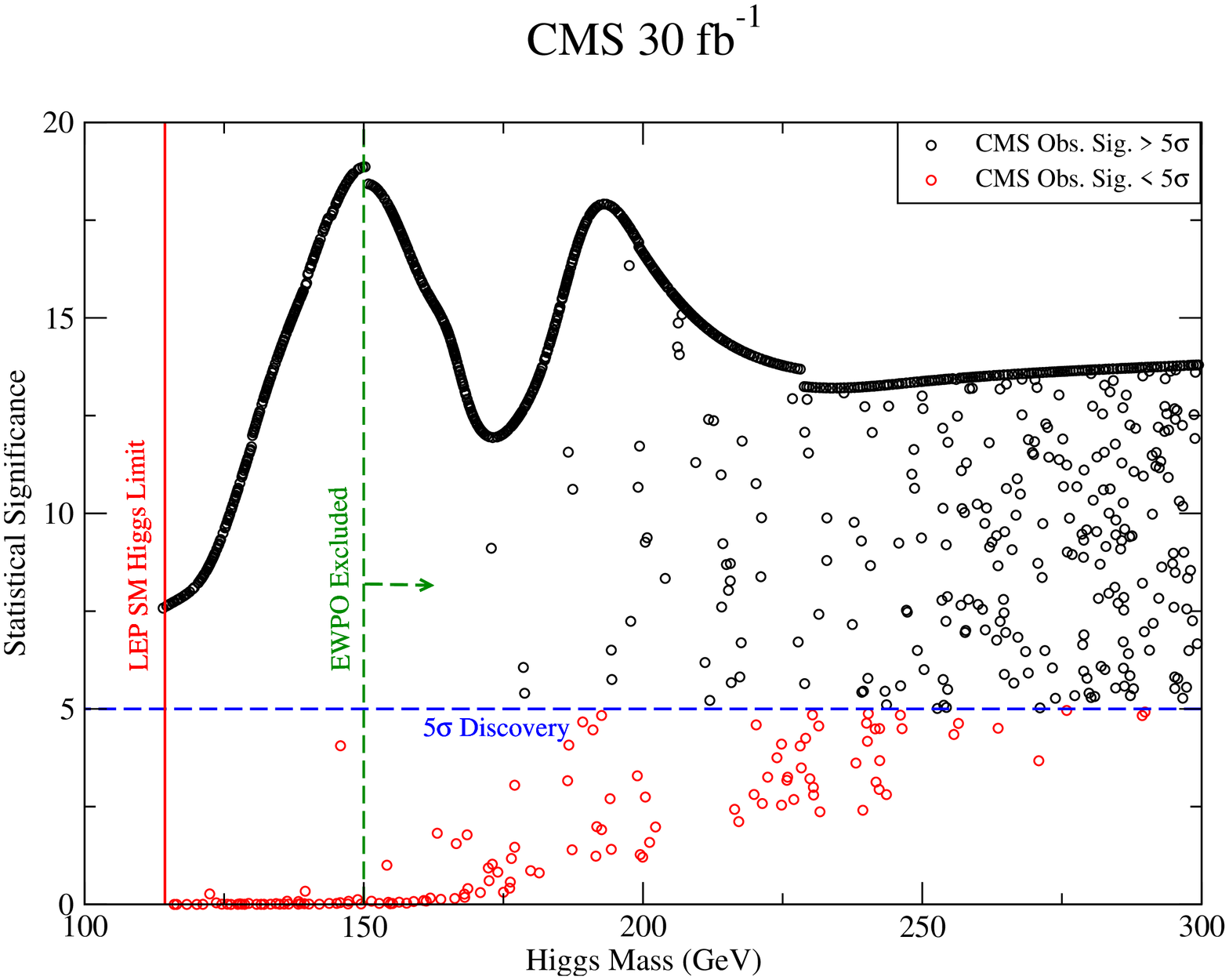}
\includegraphics[angle=-90,width=0.49\textwidth]{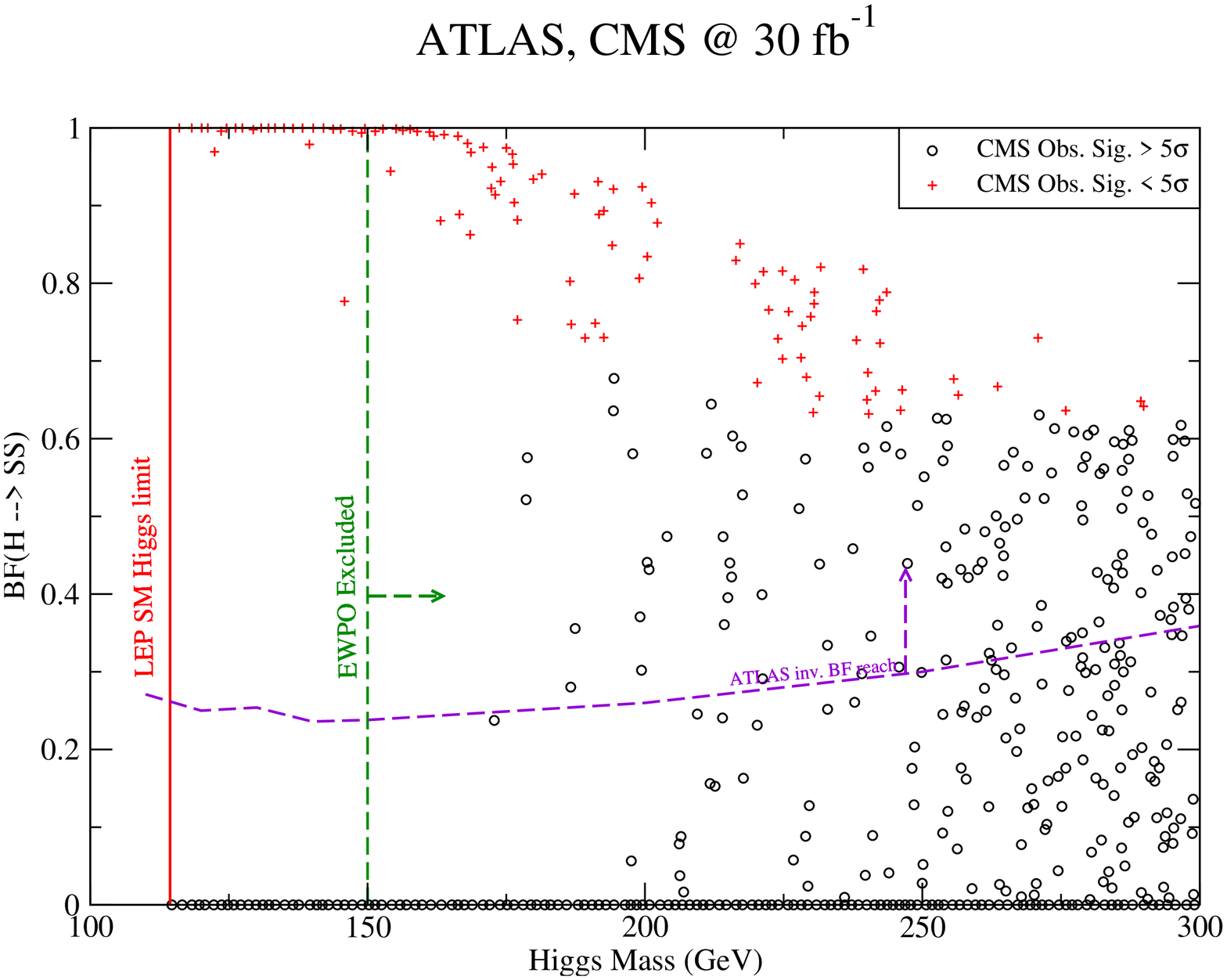}
\caption{Higgs discovery potential via direct searches at CMS with 30 fb$^{-1}$ of data (left panel) with $\Omega_S h^2 < 0.123$.  The solid curve above $5\sigma$ shows the significance expected if the decay $h\to S S$ is not kinematically accessible.  If the Higgs is allowed to decay to invisible singlet pairs, the signal from usual SM search modes is diminished.  The points that yield a lightest Higgs with a statistical significance from direct observation below $5\sigma$ are shown in red.  However, these cases are covered by the invisible search at ATLAS (right panel) via WBF \cite{atlas:2003ab}.}
\label{fig:smdm-hdisc}
\end{center}
\end{figure}

The presence of a DM-viable singlet scalar can lead to significant changes to the SM Higgs boson searches at the LHC.  If the singlet is sufficiently light enough to allow the decay $h\to SS$, the branching fractions of the Higgs boson to SM particles is reduced. Setting $H_2\to h$, $H_1\to S$, $g_{H_2}=\sin\phi=1$ in
Eq.~(\ref{eqn:bfreduction}),
\be
\text{BF}(h \to X_{SM}) =\text{BF}(h_{SM} \to X_{SM}){ \Gamma_{h_{SM}}\over  \Gamma_{h_{SM}}+\Gamma({h \to SS})},
\ee
where the decay to singlet pairs is now given by
\be
\Gamma(h\to SS) = {\delta_2^2 v^2 \over 32 \pi M_h}\sqrt{1-{4M_S^2\over M_h^2}}.
\ee
This rate can be large enough to substantially reduce the strength of the Higgs boson signal at the LHC.  Limits from LEP on the invisible decay of the Higgs boson have placed a lower bound on the SM-like Higgs boson at 114.4 GeV \cite{LEP:2001xz}.  In the left panel of Fig. \ref{fig:smdm-hdisc}, the discovery potential is shown for the SM Higgs boson at CMS with 30 fb$^{-1}$ of data \cite{ref:cmstdr} if the Higgs boson can decay to singlet pairs while $\Omega_S h^2 < 0.123$.  The solid curve above $5\sigma$ is the SM expectation of the statistical significance, while those points falling below this curve correspond to the $h\to SS$ decay being open~\footnote{The discontinuities in the statistical significance curve for the SM Higgs boson are due to the limited mass range of some search modes.  Therefore, abrupt, but small changes in the total significance occur.}.  A majority of points with a smaller significance than the SM traditional modes are correlated with large Higgs boson masses since $M_h > 2 M_S$.  

For scenarios with a light Higgs mass, however, the singlet decays easily dominate over the fermionic decay modes, leading to scenarios where Higgs discovery cannot be made.  A fraction of the points in the parameter scans have a lightest Higgs below $5\sigma$ statistical significance as shown in red.  In these cases, it is not possible to detect the $h$ boson through the usual SM processes with the given luminosity. In particular, all models consistent with the EWPO upper bound on $m_h$ and $\Omega_S h^2< 0.123$ would not be observable by CMS with 30 fb$^{-1}$ integrated luminosity. Detection may, however, be feasible by considering the Higgs decay to invisible states.  For all of the models with $< 5\sigma$ significance in CMS, the invisible branching fraction, $\text{BF}(h\to SS)$, is dominant with values ranging from 60-100\%.  This fraction of invisible decays gives observable signals at the LHC in Weak Boson Fusion (WBF) by looking for missing energy from the Higgs decay and cutting on the azimuthal correlation of the forward jets \cite{Eboli:2000ze,atlas:2003ab,Barger:2006sk}. The ATLAS invisible BF reach is illustrated in the right panel of Fig.~\ref{fig:smdm-hdisc}. An alternative method is to use $Z$-Higgstrahlung \cite{Davoudiasl:2004aj}.  We note in passing that these methods cannot work for the $H_2\to H_1 H_1$ channel in the Higgs mixing case, since the light scalars decay to SM final states and there is no missing energy. Consequently, the missing energy cut removes these events.

Another independent probe of the connection between the SM Higgs boson and the singlet DM particle is to measure the total width of the Higgs boson.  At the LHC, the Higgs width may be measured if the SM Higgs boson mass is larger than 200 GeV via the golden mode, $h\to ZZ\to 4l$ \cite{ref:atltdr}.  With 300 fb$^{-1}$ of data, the uncertainty in the Higgs width can be measured to 10\% at ATLAS for $M_h=300$ GeV.  If the Higgs decay width to singlet pairs is significant, it can alter the total Higgs width considerably.  Indeed, the lower values obtained for the statistical significance of the Higgs boson at $M_h=300$ GeV indicate that the Higgs width to singlets is nearly twice the width to SM modes~\footnote{This is dependent on the assumed scan limits.  A more liberal scan can further increase the partial width to singlet pairs.  The total SM Higgs width to SM modes at $M_h=300$ GeV is 8.5 GeV }.  However, with an increased decay rate to invisible states, the rate to $4l$ final states is decreased, and may spoil the resolution of the Higgs width measurement.  In an extreme case, the Higgs decay width can be so large that reconstructing the Higgs signal from its decay may be difficult.  

An interesting limit to consider is the case where the $\mathbb Z_2$ symmetry is not exact.  Then the decay of the Higgs boson can be quite different than either the Higgs-singlet mixing case or the DM case.  For very small Higgs-singlet mixing the light singlet-like state has a very small total decay width.  In this case, the singlet cannot be a DM candidate as it will decay long before freeze out.  However, the small mixing may lead to long decay lengths if $\phi \sim \mathcal{O}(10^{-6})$, with length scales on the order of the size of the detector.  This would provide a  Higgs decay signal where the decay products are displaced from the interaction region of the beam.  Similar to the light Higgs in section \ref{sect:higgs}, the resulting singlet decay will have a branching fraction to SM  modes equivalent to a SM Higgs of equal mass.  A variety of other models also predict displaced vertices from Higgs or other types of decays that may complicate Higgs searches \cite{Strassler:2006im,Strassler:2006ri,Arkani-Hamed:2004fb,Fairbairn:2006gg,Kang:2007xx}.

\section{Specific Model Illustrations}
\label{sect:illust}

To summarize the important aspects of the xSM, we show, in Table \ref{tab:illust}, a few representative cases to illustrate how the singlet affects Higgs searches at the LHC.  For the case of general singlet-Higgs mixing, denoted A1 and A2, we show one case where the heavy, SM-like Higgs boson dominantly decays to two light scalars (case A1).  In this case, the production of traditional Higgs decay modes through the $H_2$ is reduced by a factor of $\xi_2^2 = 0.001$ and therefore substantially reduces the statistical significance.  The light Higgs state also has a reduced rate due to its high singlet composition.  In case A2, we show that it is possible to observe both Higgs states if the mixing is sufficiently large.  The Higgs splitting mode is not kinematically accessible; therefore, the branching fractions to SM modes is not altered from the SM expectation.  With a relatively large amount of mixing, the statistical significance of both states reaches above the required level for discovery.

\begin{table}[h]
\caption{Model illustrations for the singlet-Higgs mixing case (A1,A2) and the DM, $\mathbb Z_2$, case (B1,B2).  The Higgs boson statistical significance assumes 30 fb$^{-1}$ of data from CMS.}
\begin{center}
\begin{tabular}{|c|cc|cc|}
\hline
&	A1&	A2	&B1	&B2\\
\hline
$\lambda$&0.60&0.70&0.8&0.48\\
$\delta_1$ (GeV)&3.87&-44.1&0&0\\
$\delta_2$&-1.89&2.86&-0.25&0.042\\
$\kappa_2$ (GeV)$^2$&60960&-62770&12840&1741\\
$\kappa_3$ (GeV)&318&935&0&0\\
\hline\hline
$M_{H_1}$ (GeV)&61.5&130&156&120\\
$M_{H_2}$ (GeV)&135&167&--&--\\
\hline\hline
$M_{h}$ (GeV)&--&--&156&120\\
$M_{S}$ (GeV)&--&--&72.2&55\\
\hline
$\xi^2_1$&0.001&0.62&0.25&0.47\\
$\xi^2_2$&0.001&0.38&--&--\\
\hline
$S(H_1)/\sqrt{B}$&0.00&7.1&0.1&3.8\\
$S(H_2)/\sqrt{B}$&0.01&5.2&--&--\\
\hline
$\text{BF}(H_2\to H_1 H_1)$&0.999&0.00&--&--\\
$\text{BF}(H\to SS)$&--&--&0.75&0.53\\
\hline
$\Omega_{S}h^2$&--&--&0.100&0.110\\
$\sigma_{S-p}$ (pb)&--&--&3.9$\times 10^{-8}$&5.3$\times 10^{-9}$\\
\hline
$\Delta \chi^2$&5.1&5.8&7.0&4.0\\
\hline
\end{tabular}
\end{center}
\label{tab:illust}
\end{table}%

In the case of a stable singlet, the SM Higgs may decay into the light stable scalars thereby decreasing the rate of decay to traditional SM modes.  We show for cases B1 and B2 in Table \ref{tab:illust} sample points in parameter space that are consistent with the observed relic density of DM and are below the present direct detection limits given by Xenon10.  In case B1, the SM mode production rate through the Higgs boson is about 1/4 what is expected in the SM, and therefore reduces the traditional signals below observable levels.  However, since the branching fraction to stable singlets is dominant (approximately 75\%), discovering the Higgs via the indirect, invisible BF search by ATLAS  discussed in Section \ref{sect:higgsdecay} is possible.  In case B2, the branching fraction to the singlet states is slightly weaker, yielding a $3.8\sigma$ evidence signal for the Higgs boson.  However, with increased luminosity, it may be possible to discover the Higgs boson using both direct and indirect channels in this case.

\section{Conclusions}
\label{sect:concl}

Discovering the mechanism of electroweak symmetry breaking (EWSB) is a major goal of LHC experiments. While the SM  contains a simple picture of EWSB implying the existence of one yet unseen scalar, it is entirely possible that the scalar sector is more complicated and that the larger framework that incorporates the SM contains additional light scalar degrees of freedom. Indeed, a variety of models contain such scalars, and their presence can help solve important puzzles, such as the abundance of visible and dark matter. In this paper, we have studied a simple extension of the SM -- the xSM --  that adds a real scalar gauge singlet field to the SM Higgs potential and analyzed in detail its general implications for LHC Higgs phenomenology. In doing so, we considered two scenarios: one in which the singlet scalar can mix with the SM Higgs boson, and the second in which no mixing occurs and the scalar singlet is stable. Models with mixing can give rise to a strong first order electroweak phase transition as needed for electroweak baryogenesis with a SM-like scalar having mass well above the LEP 2 direct search limit. Those in which the singlet scalar is stable provide a viable candidate for the relic abundance of cold dark matter. Thus, discovery of an augmented scalar sector at the LHC and identification of its character could have important consequences for cosmology. 

In studying the xSM, we have analyzed the implications of an augmented scalar sector for Higgs discovery potential at the LHC and outlined the possible signatures of this scenario. In brief, if the singlet is allowed to mix with the SM Higgs boson, the following can occur:

\bi
\item There are two Higgs mass eigenstates.  The ligher state, $H_1$, has the same branching fractions to SM modes as a SM Higgs boson of the same mass, but generally has a coupling strength to SM fields that is reduced by $\cos\phi$, where $\phi$ is the mixing angle.  The heavier state, $H_2$, whose coupling strength is reduced by a factor $\sin \phi$, can have the same branching fractions to SM modes as the lighter state if the decay $H_2 \to H_1 H_1$ is kinematically disallowed.  

\item Consistency with electroweak precision observations (EWPO) and LEP 2 direct search limits can significantly constrain the parameter space of the Higgs mixing scenario.  The SM-like Higgs boson ($\cos^2\phi > 0.5$) has an upper bound on its mass of 220 GeV, while the singlet-like Higgs boson can generally be heavier.  The presence of an augmented scalar sector involving singlets that mix with the neutral $SU(2)$ scalar can, thus, substantially relax the tension between EWPO and LEP 2 direct searches that applies to the SM Higgs boson. The lower mass limits on the Higgs states are also affected by the singlet-Higgs mixing.  If the light Higgs boson has a large singlet fraction, its mass can be substantially lower than the LEP bound on the SM Higgs boson.  

\item Discovery of the Higgs bosons in this model is viable, but could occur under different conditions than expected for a SM Higgs. If the lighter scalar is $SU(2)$-like, then discovery at CMS with a luminosity of 30 fb$^{-1}$ is possible if its mass is less than about 180 GeV. If the heavier scalar is $SU(2)$-like, then its mass could be as large as $\sim 220$ GeV. Either way, observation of a Higgs scalar with mass above 150 GeV would point toward an augmented scalar sector.

\item If a Higgs scalar is discovered at the LHC, it may be possible to determine that its couplings to SM particles are reduced by up to $\sim 30\%$ ($\xi^2\gtrsim 0.5$). Observation of a Higgs with reduced couplings would also indicate an augmented scalar sector that includes Higgs mixing.

\item An additional signature of an augmented scalar sector is the presence of the Higgs splitting channel, $H_2\to H_1 H_1$. If this channel is kinematically allowed, a non-standard final state containing $4b$ or $2b+2\tau$ can occur. The resulting reduction in BF to conventional SM Higgs final states can be large enough to preclude discovery of the $H_2$  in these channels.  For an $SU(2)$-like $H_2$, direct observation of non-standard final states may be feasible at the LHC based on analyses specific to extended SUSY models where the Higgs phenomenology is similar, whereas identification of a singlet-like $H_2$ in this way appears challenging at best. More generally, if the Higgs splitting decays are absent, it is likely that at least one of the Higgs states can be discovered at CMS with a luminosity of 30 fb$^{-1}$.  In the strongly mixed case, it is possible that both states can be discovered.
\ei

\noindent If, however, the SM Higgs state is forbidden to mix with the singlet state by a ${\mathbb Z}_2$ symmetry of the scalar potential, the singlet is a viable dark matter candidate.  In this case, we observe the following:

\bi
\item Consistency with the observed relic DM tightly constrains the model.  Since the singlet annihilations occur only through the presence of SM Higgs boson, the annihilation rate is controlled by the Higgs-singlet coupling, $\delta_2$, the Higgs self coupling, $\lambda$, and the singlet mass (c.f. Eq. (\ref{eqn:hpot})).  

\item  Elastic scattering rates off nuclear targets are in the sensitivity range of present and future direct detection experiments.  Requiring consistency with relic density measurements and EWPO places the expected scattering rates within or near the reach of the SuperCDMS experiment.  However, in a non-thermal or other non-standard cosmological scenario, the relic density can be enhanced or suppressed by low reheating from the value obtained by the standard calculation.  In that case, a very small Higgs-singlet coupling is allowed and  the direct detection rate can be below future experimental limits.  

\item The stable singlet can have a significant impact on the search for the Higgs boson at the LHC.  If the Higgs decay to singlet pairs is kinematically allowed, the decay rates to usual SM modes are reduced.  This can lower the statistical significance of the Higgs boson signal below the $5\sigma$ excess required for discovery.  If the decay rate to stable singlets is large, the Higgs may still be discovered through its decays to invisible states via weak boson fusion. Specifically, the Higgs boson in all EWPO-allowed models that do not over produce the relic density  could only be discovered through the invisible decays. More generally, over the whole parameter range, the SM Higgs boson can be discovered by the traditional direct search and/or the indirect search. 

\item If the ${\mathbb Z}_2$ symmetry forbidding Higgs-singlet mixing is only approximate, the singlet dominated state can have a long lifetime.  For mixing angles ${\cal O}(10^{-6})$, the proper decay length of the singlet state can be the scale of the LHC detectors.  In such a scenario, this state behaves as a Higgs boson but decays with displaced vertices.  
\ei

\section{Acknowledgments}
We thank H. Baer, D. Cline and W-Y. Keung for helpful discussions.  We also thank ShinÕichiro Ando and Sean Tulin for technical discussions related to the scalar loop calculations and J. Kile and J. Erler for assistance with the GAAP code.  This work was supported in part by the U.S.~Department of Energy under grants No. DE-FG02-95ER40896 and DE-FG02-05ER41361, by the Wisconsin Alumni Research Foundation, by the Friends of the IAS, and by the National Science Foundation grants No. PHY-0503584 and PHY-PHY-0555674.

\appendix
\section{Higgs corrections to gauge boson propagators}
\label{apx:prop}
Here, we provide explicit expressions for the unrenormalized $\Pi_{VV}(q^2)$, computed in Feynman gauge using dimensional regularization in $4-2 \epsilon$ dimensions from the graphs of Fig. \ref{fig:ewfd}. The addition of a single, neutral singlet scalar only affects the $W$- and $Z$-boson propagator functions, which we provide below. Both the divergent ($1/\epsilon$) and finite parts are included:
\begin{eqnarray}
\Pi_{WW}(q^2) & = & -\frac{g^2}{2(4\pi)^2}\, \sum_{j=1-2,\, a=1-3} \left(V_H^{ja}\right)\Bigl\{-\frac{1}{6} q^2(\alpha_\epsilon+1)
+q^2 F(M_W^2,m_a^2,q^2)\\
\nonumber
&-&M_W^2 F_1(m_a^2,M_W^2,q^2)-m_a^2 F_1(M_W^2,m_a^2,q^2)
+\frac{1}{2}\left[m_a^2\ln m_a^2+ M_W^2\ln M_W^2\right]\Bigr\}\\
\nonumber
&+&\frac{g^{\prime\, 2}}{(4\pi)^2} M_W^2\Bigl\{s_W^2\left[\alpha_\epsilon-F_0(M_Z^2,M_W^2,q^2)\right]
+c_W^2\left[\alpha_\epsilon-F_0(M_Z^2,M_W^2,q^2)\right]\Bigr\}\\
\nonumber
&+&\frac{g^2}{(4\pi)^2} M_W^2 \sum_{a=1-3} \left(V_H^{1a}\right)^2\left[\alpha_\epsilon-F_0(M_W^2, m_a^2, q^2)\right]
\end{eqnarray}
\begin{eqnarray}
\Pi_{ZZ}(q^2) & = & -\frac{g^2}{2(4\pi)^2c_W^2}\, \sum_{a,b=1-3} \Bigl[(V_H^{1a})^2(V_H^{2b})^2-(V_H^{1a}V_H^{2a})(V_H^{1b}V_H^{2b})\Bigr]  \\
\nonumber 
&&\times \Bigl\{-\frac{1}{6} q^2(\alpha_\epsilon+1)
+q^2 F(m_a^2,m_b^2,q^2)\\
\nonumber
&-&m_a^2 F_1(m_b^2,m_a^2,q^2)-m_b^2 F_1(m_a^2,m_b^2,q^2)
+\frac{1}{2}\left[m_a^2\ln m_a^2+ m_b^2\ln m_b^2\right]\Bigr\}\\
\nonumber
&-&\frac{g^2}{2(4\pi)^2c_W^2}\,\left(c_W^2-s_W^2\right)^2\, \Bigl\{-\frac{1}{6} q^2(\alpha_\epsilon+1)
+q^2 F(m_W^2,m_W^2,q^2)\\
\nonumber
&-& 2M_W^2 F_1(M_W^2,M_W^2,q^2)+M_W^2\ln M_W^2\Bigr\}\\
\nonumber
&+&2\frac{g^{\prime\, 2} s_W^2}{(4\pi)^2} M_W^2\left[\alpha_\epsilon-F_0(M_W^2,M_W^2,q^2)\right]\\
\nonumber
&+& \frac{(g^2+g^{\prime\, 2})}{(4\pi)^2} M_Z^2\, \sum_{a=1-3}\left(V_H^{1a}\right)^2\, \left[\alpha_\epsilon-F_0(M_Z^2,m_a^2,q^2)\right]\ \ \ ,
\end{eqnarray}
where
\be
\alpha_\epsilon=\frac{1}{\epsilon}-\gamma+\ln 4\pi\mu^2
\ee
with $\gamma$ being Euler's constant, $\mu$ the 't Hooft scale, and $F\equiv F_1-F_2$. In deriving these expressions, we have used $M_W$ and $M_Z$ as the masses appearing in the charged and neutral would-be Goldstone boson propagators as dictated by the Feynman gauge. The $\overline{MS}$ renormalized propagator functions ${\hat\Pi}_{VV}(q^2)$ are obtained by subtracting the $1/\epsilon-\gamma+\ln 4\pi$ and choosing an appropriate value for $\mu$, which we take to be $M_Z$. These expressions reduce to those appearing in Ref.~\cite{Degrassi:1993kn} in the Standard Model case: $V_H^{11}=V_H^{23}$ and all other $V_H^{ij}=0$. As indicated in the main text, the foregoing expressions can apply to the case of one complex $SU(2)$ doublet and $N$ real scalars by expanding the dimension of the matrix $V_H^{ij}$.

\bibliographystyle{h-physrev}
\bibliography{ssm-pheno9}                      

\begin{thebibliography}{100}

\bibitem{Barate:2003sz}
LEP Working Group for Higgs boson searches, R.~Barate {\em et~al.},
\newblock Phys. Lett. {\bf B565}, 61 (2003), hep-ex/0306033.

\bibitem{McDonald:1993ex}
J.~McDonald,
\newblock Phys. Rev. {\bf D50}, 3637 (1994).

\bibitem{Bento:2000ah}
M.~C. Bento, O.~Bertolami, R.~Rosenfeld, and L.~Teodoro,
\newblock Phys. Rev. {\bf D62}, 041302 (2000), astro-ph/0003350.

\bibitem{Burgess:2000yq}
C.~P. Burgess, M.~Pospelov, and T.~ter Veldhuis,
\newblock Nucl. Phys. {\bf B619}, 709 (2001), hep-ph/0011335.

\bibitem{Davoudiasl:2004be}
H.~Davoudiasl, R.~Kitano, T.~Li, and H.~Murayama,
\newblock Phys. Lett. {\bf B609}, 117 (2005), hep-ph/0405097.

\bibitem{Schabinger:2005ei}
R.~Schabinger and J.~D. Wells,
\newblock Phys. Rev. {\bf D72}, 093007 (2005), hep-ph/0509209.

\bibitem{O'Connell:2006wi}
D.~O'Connell, M.~J. Ramsey-Musolf, and M.~B. Wise,
\newblock (2006), hep-ph/0611014.

\bibitem{Kusenko:2006rh}
A.~Kusenko,
\newblock Phys. Rev. Lett. {\bf 97}, 241301 (2006), hep-ph/0609081.

\bibitem{Bahat-Treidel:2006kx}
O.~Bahat-Treidel, Y.~Grossman, and Y.~Rozen,
\newblock (2006), hep-ph/0611162.

\bibitem{Barger:2006dh}
V.~Barger, P.~Langacker, H.-S. Lee, and G.~Shaughnessy,
\newblock Phys. Rev. {\bf D73}, 115010 (2006), hep-ph/0603247.

\bibitem{Barger:2006sk}
V.~Barger, P.~Langacker, and G.~Shaughnessy,
\newblock Phys. Rev. {\bf D75}, 055013 (2007), hep-ph/0611239.

\bibitem{Barger:2007nv}
V.~Barger {\em et~al.},
\newblock Phys. Rev. {\bf D75}, 115002 (2007), hep-ph/0702036.

\bibitem{Barger:2007ay}
V.~Barger, P.~Langacker, and G.~Shaughnessy,
\newblock (2007), hep-ph/0702001.

\bibitem{Ellis:1988er}
J.~R. Ellis, J.~F. Gunion, H.~E. Haber, L.~Roszkowski, and F.~Zwirner,
\newblock Phys. Rev. {\bf D39}, 844 (1989).

\bibitem{NMSSM1}
M.~Bastero-Gil, C.~Hugonie, S.~F. King, D.~P. Roy, and S.~Vempati,
\newblock Phys. Lett. {\bf B489}, 359 (2000), hep-ph/0006198.

\bibitem{NMSSM2}
A.~de~Gouvea, A.~Friedland, and H.~Murayama,
\newblock Phys. Rev. {\bf D57}, 5676 (1998), hep-ph/9711264.

\bibitem{nMSSM}
C.~Panagiotakopoulos and K.~Tamvakis,
\newblock Phys. Lett. {\bf B469}, 145 (1999), hep-ph/9908351.

\bibitem{Panagiotakopoulos:2000wp}
C.~Panagiotakopoulos and A.~Pilaftsis,
\newblock Phys. Rev. {\bf D63}, 055003 (2001), hep-ph/0008268.

\bibitem{Menon:2004wv}
A.~Menon, D.~E. Morrissey, and C.~E.~M. Wagner,
\newblock Phys. Rev. {\bf D70}, 035005 (2004), hep-ph/0404184.

\bibitem{Dedes:2000jp}
A.~Dedes, C.~Hugonie, S.~Moretti, and K.~Tamvakis,
\newblock Phys. Rev. {\bf D63}, 055009 (2001), hep-ph/0009125.

\bibitem{UMSSM1}
M.~Cvetic, D.~A. Demir, J.~R. Espinosa, L.~L. Everett, and P.~Langacker,
\newblock Phys. Rev. {\bf D56}, 2861 (1997), hep-ph/9703317.

\bibitem{umssm2}
P.~Langacker and J.~Wang,
\newblock Phys. Rev. {\bf D58}, 115010 (1998), hep-ph/9804428.

\bibitem{deCarlos:1997yv}
B.~de~Carlos and J.~R. Espinosa,
\newblock Phys. Lett. {\bf B407}, 12 (1997), hep-ph/9705315.

\bibitem{smssm}
J.~Erler, P.~Langacker, and T.-j. Li,
\newblock Phys. Rev. {\bf D66}, 015002 (2002), hep-ph/0205001.

\bibitem{Han:2004yd}
T.~Han, P.~Langacker, and B.~McElrath,
\newblock Phys. Rev. {\bf D70}, 115006 (2004), hep-ph/0405244.

\bibitem{Choi:2006fz}
S.~Y. Choi, H.~E. Haber, J.~Kalinowski, and P.~M. Zerwas,
\newblock (2006), hep-ph/0612218.

\bibitem{Cheung:2007sv}
K.~Cheung, J.~Song, and Q.-S. Yan,
\newblock (2007), hep-ph/0703149.

\bibitem{King:2005my}
S.~F. King, S.~Moretti, and R.~Nevzorov,
\newblock Phys. Lett. {\bf B634}, 278 (2006), hep-ph/0511256.

\bibitem{King:2005jy}
S.~F. King, S.~Moretti, and R.~Nevzorov,
\newblock Phys. Rev. {\bf D73}, 035009 (2006), hep-ph/0510419.

\bibitem{Li:2006xb}
T.-j. Li,
\newblock (2006), hep-ph/0612359.

\bibitem{Lopez-Fogliani:2005yw}
D.~E. Lopez-Fogliani and C.~Munoz,
\newblock Phys. Rev. Lett. {\bf 97}, 041801 (2006), hep-ph/0508297.

\bibitem{Schuster:2005py}
P.~C. Schuster and N.~Toro,
\newblock (2005), hep-ph/0512189.

\bibitem{Profumo:2007wc}
S.~Profumo, M.~J. Ramsey-Musolf, and G.~Shaughnessy,
\newblock (2007), arXiv:0705.2425 [hep-ph].

\bibitem{Sikivie:2006ni}
P.~Sikivie,
\newblock (2006), astro-ph/0610440.

\bibitem{Kile:2007ts}
J.~Kile and M.~J. Ramsey-Musolf,
\newblock (2007), arXiv:0705.0554 [hep-ph].

\bibitem{LEPEWWG:2007}
M.~W. GrŸnewald {\em et~al.},
\newblock http://lepewwg.web.cern.ch/LEPEWWG/plots/winter2007/.

\bibitem{Peskin:2001rw}
M.~E. Peskin and J.~D. Wells,
\newblock Phys. Rev. {\bf D64}, 093003 (2001), hep-ph/0101342.

\bibitem{Spergel:2006hy}
D.~N. Spergel {\em et~al.},
\newblock (2006), astro-ph/0603449.

\bibitem{Yao:2006px}
Particle Data Group, W.~M. Yao {\em et~al.},
\newblock J. Phys. {\bf G33}, 1 (2006).

\bibitem{Eboli:2000ze}
O.~J.~P. Eboli and D.~Zeppenfeld,
\newblock Phys. Lett. {\bf B495}, 147 (2000), hep-ph/0009158.

\bibitem{Davoudiasl:2004aj}
H.~Davoudiasl, T.~Han, and H.~E. Logan,
\newblock Phys. Rev. {\bf D71}, 115007 (2005).

\bibitem{DMSAG}
H.~Sobel {\em et~al.},
\newblock in press.

\bibitem{Kribs:2001ic}
G.~D. Kribs,
\newblock (2001), hep-ph/0110242.

\bibitem{Dermisek:2006wr}
R.~Dermisek and J.~F. Gunion,
\newblock Phys. Rev. {\bf D75}, 075019 (2007), hep-ph/0611142.

\bibitem{Dermisek:2005gg}
R.~Dermisek and J.~F. Gunion,
\newblock Phys. Rev. {\bf D73}, 111701 (2006), hep-ph/0510322.

\bibitem{Arnold:1992rz}
P.~Arnold and O.~Espinosa,
\newblock Phys. Rev. {\bf D47}, 3546 (1993), hep-ph/9212235.

\bibitem{Dermisek:2006py}
R.~Dermisek, J.~F. Gunion, and B.~McElrath,
\newblock (2006), hep-ph/0612031.

\bibitem{Wood:1997zq}
C.~S. Wood {\em et~al.},
\newblock Science {\bf 275}, 1759 (1997).

\bibitem{Anthony:2003ub}
SLAC E158, P.~L. Anthony {\em et~al.},
\newblock Phys. Rev. Lett. {\bf 92}, 181602 (2004), hep-ex/0312035.

\bibitem{Erler:1999ug}
J.~Erler,
\newblock (1999), hep-ph/0005084.

\bibitem{cdfcollab:2007bx}
CDF,
\newblock (2007), hep-ex/0703034.

\bibitem{Casas:1996aq}
J.~A. Casas, J.~R. Espinosa, and M.~Quiros,
\newblock Phys. Lett. {\bf B382}, 374 (1996), hep-ph/9603227.

\bibitem{Casas:1994qy}
J.~A. Casas, J.~R. Espinosa, and M.~Quiros,
\newblock Phys. Lett. {\bf B342}, 171 (1995), hep-ph/9409458.

\bibitem{Hambye:1997ax}
T.~Hambye and K.~Riesselmann,
\newblock (1997), hep-ph/9708416.

\bibitem{Hambye:1996wb}
T.~Hambye and K.~Riesselmann,
\newblock Phys. Rev. {\bf D55}, 7255 (1997), hep-ph/9610272.

\bibitem{Altarelli:1994rb}
G.~Altarelli and G.~Isidori,
\newblock Phys. Lett. {\bf B337}, 141 (1994).

\bibitem{Ford:1992mv}
C.~Ford, D.~R.~T. Jones, P.~W. Stephenson, and M.~B. Einhorn,
\newblock Nucl. Phys. {\bf B395}, 17 (1993), hep-lat/9210033.

\bibitem{Sher:1988mj}
M.~Sher,
\newblock Phys. Rept. {\bf 179}, 273 (1989).

\bibitem{Lindner:1988ww}
M.~Lindner, M.~Sher, and H.~W. Zaglauer,
\newblock Phys. Lett. {\bf B228}, 139 (1989).

\bibitem{Lindner:1985uk}
M.~Lindner,
\newblock Zeit. Phys. {\bf C31}, 295 (1986).

\bibitem{us:2007xx}
V.~Barger, P.~Langacker, M.~Ramsey-Musolf, and G.~Shaughnessy,
\newblock in preparation.

\bibitem{ref:cmstdr}
A.~DeRoeck,
\newblock CERN/LHCC 2006-021.

\bibitem{ref:atltdr}
A.~Collaboration,
\newblock CERN/LHCC 99-15.

\bibitem{Duhrssen:2004uu}
M.~Duhrssen {\em et~al.},
\newblock (2004), hep-ph/0407190.

\bibitem{Zeppenfeld:2000td}
D.~Zeppenfeld, R.~Kinnunen, A.~Nikitenko, and E.~Richter-Was,
\newblock Phys. Rev. {\bf D62}, 013009 (2000), hep-ph/0002036.

\bibitem{Abe:2001np}
American Linear Collider Working Group, T.~Abe {\em et~al.},
\newblock (2001), hep-ex/0106056.

\bibitem{Dermisek:2005ar}
R.~Dermisek and J.~F. Gunion,
\newblock Phys. Rev. Lett. {\bf 95}, 041801 (2005), hep-ph/0502105.

\bibitem{Carena:2007xx}
M.~Carena, T.~Han, G.~Huang, and C.~Wagner,
\newblock in preparation.

\bibitem{Espinosa:2007qk}
J.~R. Espinosa and M.~Quiros,
\newblock (2007), hep-ph/0701145.

\bibitem{Kolb:1990vq}
E.~W. Kolb and M.~S. Turner,
\newblock Front. Phys. {\bf 69}, 1 (1990).

\bibitem{Gondolo:2004sc}
P.~Gondolo {\em et~al.},
\newblock JCAP {\bf 0407}, 008 (2004), astro-ph/0406204.

\bibitem{Salati:2002md}
P.~Salati,
\newblock Phys. Lett. {\bf B571}, 121 (2003), astro-ph/0207396.

\bibitem{Profumo:2003hq}
S.~Profumo and P.~Ullio,
\newblock JCAP {\bf 0311}, 006 (2003), hep-ph/0309220.

\bibitem{Rosati:2003yw}
F.~Rosati,
\newblock Phys. Lett. {\bf B570}, 5 (2003), hep-ph/0302159.

\bibitem{Chung:2007vz}
D.~J.~H. Chung, L.~L. Everett, and K.~T. Matchev,
\newblock (2007), arXiv:0704.3285.

\bibitem{Baltz:2006fm}
E.~A. Baltz, M.~Battaglia, M.~E. Peskin, and T.~Wizansky,
\newblock Phys. Rev. {\bf D74}, 103521 (2006).

\bibitem{Carena:2006nv}
M.~Carena, D.~Hooper, and A.~Vallinotto,
\newblock Phys. Rev. {\bf D75}, 055010 (2007).

\bibitem{Seidel:2005kx}
W.~Seidel,
\newblock Nucl. Phys. Proc. Suppl. {\bf 138}, 130 (2005).

\bibitem{Morales:2005ky}
A.~Morales,
\newblock Nucl. Phys. Proc. Suppl. {\bf 138}, 135 (2005).

\bibitem{Akerib:2006rr}
D.~S. Akerib {\em et~al.},
\newblock Nucl. Instrum. Meth. {\bf A559}, 411 (2006).

\bibitem{Akerib:2005za}
CDMS, D.~S. Akerib {\em et~al.},
\newblock Phys. Rev. {\bf D73}, 011102 (2006), astro-ph/0509269.

\bibitem{Akerib:2006ri}
D.~S. Akerib {\em et~al.},
\newblock Nucl. Instrum. Meth. {\bf A559}, 390 (2006).

\bibitem{Akerib:2005kh}
D.~S. Akerib {\em et~al.},
\newblock Phys. Rev. Lett. {\bf 96}, 011302 (2006).

\bibitem{Sanglard:2005we}
The EDELWEISS, V.~Sanglard {\em et~al.},
\newblock Phys. Rev. {\bf D71}, 122002 (2005), astro-ph/0503265.

\bibitem{Benetti:2007cd}
P.~Benetti {\em et~al.},
\newblock (2007), astro-ph/0701286.

\bibitem{Angle:2007uj}
J.~Angle {\em et~al.},
\newblock (2007), arxiv:0706.0039.

\bibitem{Ni:2006fh}
K.~Ni and L.~Baudis,
\newblock (2006), astro-ph/0611124.

\bibitem{WARP}
R.~Brunetti {\em et~al.},
\newblock (2004), http://warp.lngs.infn.it.

\bibitem{Aprile:2002ef}
E.~Aprile {\em et~al.},
\newblock (2002), astro-ph/0207670.

\bibitem{Bolte:2006pf}
W.~J. Bolte {\em et~al.},
\newblock J. Phys. Conf. Ser. {\bf 39}, 126 (2006).

\bibitem{Aubin:2006rc}
F.~Aubin {\em et~al.},
\newblock AIP Conf. Proc. {\bf 828}, 265 (2006).

\bibitem{Ellis:2000ds}
J.~R. Ellis, A.~Ferstl, and K.~A. Olive,
\newblock Phys. Lett. {\bf B481}, 304 (2000), hep-ph/0001005.

\bibitem{Giudice:2000ex}
G.~F. Giudice, E.~W. Kolb, and A.~Riotto,
\newblock Phys. Rev. {\bf D64}, 023508 (2001), hep-ph/0005123.

\bibitem{LEP:2001xz}
LEP Higgs Working for Higgs boson searches,
\newblock (2001), hep-ex/0107032.

\bibitem{atlas:2003ab}
L.~Neukermans and B.~Girolamo,
\newblock (2003), ATL-PHYS-2003-006.

\bibitem{Strassler:2006im}
M.~J. Strassler and K.~M. Zurek,
\newblock (2006), hep-ph/0604261.

\bibitem{Strassler:2006ri}
M.~J. Strassler and K.~M. Zurek,
\newblock (2006), hep-ph/0605193.

\bibitem{Arkani-Hamed:2004fb}
N.~Arkani-Hamed and S.~Dimopoulos,
\newblock JHEP {\bf 06}, 073 (2005), hep-th/0405159.

\bibitem{Fairbairn:2006gg}
M.~Fairbairn {\em et~al.},
\newblock Phys. Rept. {\bf 438}, 1 (2007), hep-ph/0611040.

\bibitem{Kang:2007xx}
J.~Kang, P.~Langacker, and B.~Nelson,
\newblock in preparation.

\bibitem{Degrassi:1993kn}
G.~Degrassi, B.~A. Kniehl, and A.~Sirlin,
\newblock Phys. Rev. {\bf D48}, 3963 (1993).

\end{thebibliography}

\newpage

\end{document}